\documentclass[prx,twocolumn,longbibliography,superscriptaddress,nofootinbib]{revtex4-1}

\usepackage{graphicx}
\usepackage{amsmath,mathrsfs}
\usepackage{amsfonts}
\usepackage{amssymb}
\usepackage{wasysym}
\usepackage{dsfont}
\usepackage{color}

\definecolor{linkcolor}{rgb}{0,0.1,0.7}
\usepackage[colorlinks=true,allcolors=linkcolor]{hyperref}

\renewcommand{\vec}[1]{{\boldsymbol{\mathbf{#1}}}}
\newcommand{\llangle}{\langle\!\langle}
\newcommand{\rrangle}{\rangle\!\rangle}

\begin{document}

\title{
Competition for Stem Cell Fate Determinants \\
as a Mechanism for Tissue Homeostasis
}

\author{David J. J\"org}
\affiliation{Cavendish Laboratory, Department of Physics, University of Cambridge, JJ Thomson Avenue, Cambridge CB3 0HE, United Kingdom}
\affiliation{The Wellcome Trust/Cancer Research UK Gurdon Institute, University of Cambridge, Tennis Court Road, Cambridge CB2 1QN, United Kingdom}

\author{Yu Kitadate}
\author{Shosei Yoshida}
\affiliation{Division of Germ Cell Biology, National Institute for Basic Biology, National Institutes of Natural Sciences, 5-1 Higashiyama, Myodaiji, Okazaki 444-8787, Japan}
\affiliation{Department of Basic Biology, School of Life Science, Graduate University for Advanced Studies (Sokendai), 5-1 Higashiyama, Myodaiji, Okazaki 444-8787, Japan}

\author{Benjamin D. Simons}\email{bds10@cam.ac.uk}
\affiliation{Cavendish Laboratory, Department of Physics, University of Cambridge, JJ Thomson Avenue, Cambridge CB3 0HE, United Kingdom}
\affiliation{The Wellcome Trust/Cancer Research UK Gurdon Institute, University of Cambridge, Tennis Court Road, Cambridge CB2 1QN, United Kingdom}
\affiliation{The Wellcome Trust/Medical Research Council Stem Cell Institute, University of Cambridge, Cambridge CB2 1QR, United Kingdom}

\begin{abstract}

\noindent%
Stem cells maintain tissues by generating differentiated cell types while simultaneously self-renewing their own population.
The mechanisms that allow stem cell populations to function collectively to control their density, maintain robust homeostasis and recover from injury remain elusive.
Motivated by recent experimental advances, here we develop a generic and robust mechanism of stem cell self-renewal based on competition for diffusible fate determinants, which exert fate control.
Using both analytical methods and stochastic simulations, we show that the mechanism is characterized by signature dynamic and statistical properties, from stem cell density fluctuations and transient large-scale oscillation dynamics during recovery, to scaling clonal dynamics, front-like boundary propagation and features of dynamical `quantization' of self-renewal zones in localized domains.
Based on these findings, we suggest that competition for fate determinants provides a generic and robust mechanism by which stem cells can self-organize to achieve density homeostasis in an open (or `facultative') niche environment.
\end{abstract}

\maketitle
 
\section{Introduction}
\noindent%
A hallmark of living systems is their ability to renew. In multicellular organisms, this renewal capacity is invested in the germ line across generations, and through  the long-term renewal capacity of its constituent tissues during the individual's life span. The ability of cycling adult tissues, such as the skin epidermis, gut epithelium and blood, to renew and repair relies on the activity of resident adult stem cells, which divide and differentiate to replenish functional cells lost through homeostatic turnover, while maintaining their pool size \cite{Siminovitch1963}.
To achieve long-term homeostasis, the stem cell population must tightly balance duplication and differentiation to ensure that tissues are maintained with the correct size, pattern and cell composition \cite{Simons2011}. Resolving the mechanisms by which stem cells achieve perfect self-renewal represents a defining question in tissue stem cell biology.

To resolve patterns of stem cell density regulation, emphasis has been placed on cell lineage tracing assays.
Measurements based on the activation of reporter proteins in targeted cell populations using transgenic animal models \cite{Clayton2007,Barker2007,Nakagawa2007,Klein2010,Snippert2010,LopezGarcia2010,Driessens2012,Kretzschmar2012,Schepers2012,Blanpain2013,Rulands2018} and genetic barcoding \cite{Schepers2008,Gerrits2010,Lu2011,Verovskaya2013,Nguyen2014,Thielecke2017,Lan2017}, as well as intravital live-imaging assays \cite{Nakagawa2010,Rompolas2012,Hara2014,Ritsma2014,Pilz2018,Mesa2018}, have allowed the fate of marked cells and the differentiating progeny---termed `clones'---to be traced over time.
From the quantitative analysis of clone size distributions, rules of cell fate decision making have begun to emerge.
Applied to cycling adult tissues, these studies have shown that the balance between stem cell duplication and differentiation is, in many cases, not enforced at the level of individual cell fate decisions---in the manner of invariant asymmetric fate---but apply only at the level of the population \cite{Clayton2007,Nakagawa2007,LopezGarcia2010,Snippert2010,Klein2010,Simons2011,Rulands2017}.
In this process of `population asymmetric' self-renewal, the outcome of individual stem cell divisions may be probabilistic, with fate acquired according to defined statistical dependencies \cite{Loeffler2002,Klein2010}.
But what is the regulatory basis of stochastic cell fate decision-making?

To address the dynamics of stem cell fate, early studies placed emphasis on finding intrinsic mechanisms, in which the balance between cell duplication and differentiation is enforced through cell-autonomous regulatory programs.
However, the remarkable ability of adult stem cell populations to both maintain and repair tissues following injury suggests mechanisms of fate control that are sensitive to signals from neighboring cells as well as local environmental cues which together constitute the stem cell `niche'.
In some cases, such as the mammalian intestinal `crypt' and \textit{Drosophila} testis, the niche finds an anatomical basis in the spatial organization of specialized somatic cells to which stem cells associate. However, in other cases, such as mammalian hematopoiesis in the bone marrow, stem cells, which form a scattered minority population, actively migrate, roaming among their differentiating progeny---an organization sometimes known as an `open' (or `facultative') niche~\cite{Yoshida2007,Morrison2008,Stine2013,Hara2014,Yoshida2018}. 

In tissues supported by a `closed' anatomical niche, competition for limited niche access provides a simple framework to regulate density homeostasis: As stem cells duplicate, only one half of their progeny can retain niche access while the other half becomes displaced, loosing physical contact with the niche and entering into a differentiation program \cite{Simons2011}. However, in systems supported by an open niche, the mechanisms that ensure density regulation during homeostasis and following injury are less obvious: How can stem cells, widely separated from neighboring stem cells by multiple cell diameters, sense their local density and adjust their fate to maintain homeostasis?

The situation is exemplified by spermatogonial stem cell regulation. In the mouse, sperm production relies on the activity of stem cells that divide and differentiate, giving rise to progeny that expand through serial rounds of cell division, before entering into a maturation step involving meiosis and translocation towards the lumen of the tubules.
Recently, combining marker-based assays with studies of stem and progenitor cell kinetics both in homeostasis and during recovery following injury, we have found evidence that the regulation of stem cell homeostasis involves the competition for a limited supply of diffusible factors released from lymphatic endothelial cells that contribute to the testicular open niche environment \cite{Kitadate2019}.
These fate determinants are thought to exert a regulatory control on stem cells, `priming' their fate towards either duplication or loss through differentiation in a concentration-dependent manner. In contrast to analogous mechanisms based on competition for nutritional resources, as often considered in ecological systems \cite{Klausmeier2002,Amarasekare2003,Kim2016}, these fate determinants do not regulate population size by imposing a limited energy supply that leads to the starvation of excess populations.
Instead, this mechanism effectively provides a feedback of the cell density on stem cell fate through a regulatory fate-steering process.

Although conceived as a mechanism of stem cell self-renewal during murine spermatogenesis, competition for fate determinants provides a generic framework for stem cell density regulation with potential applications to other tissues supported by an open niche environment.
Therefore, based on this hypothesis, here we introduce and investigate a general individual-based model of stem cell density regulation based on motile stem cells interacting with a diffusible fate determinant through consumption and fate control.
Recent theoretical studies of motile `communities' have shown that interesting collective phenomena can arise if their dynamics depends on the presence of neighbors; these include the emergence of patterned states, clustering phenomena and dynamical crossovers between scaling regimes \cite{Fuentes2003,Birch2006,Brigatti2008,HernandezGarcia2004,Ramos2008,Heinsalu2010,Heinsalu2012,HernandezGarcia2015,Yamaguchi2017}.
Here we investigate the conditions to achieve robust stem cell density maintenance, both in homeostasis and in recovery following perturbations away from steady state (Section~\ref{sec:homeostasis}); the statistical properties of the homeostatic state such as stem cell density fluctuations (Section~\ref{sec:fluctuations}); and the hallmark statistical scaling properties of individually labelled cell clones (Section~\ref{sec:clonal.dynamics}).
The presence of a dynamic concentration field of fate determinants enables a study of the dependence of these dynamic features on the kinetics of the fate determinant.
Moreover, we systematically study the recovery behavior when the system is driven far away from its steady state such as during initial colonization scenarios (Section~\ref{sec:colonization}) and the behavior of `mutated' subpopulations with a proliferative advantage (Section~\ref{sec:advantage}).
Finally, we show that the system can give rise to peculiar quantization phenomena and how they emerge from the dynamics of motility and self-renewal (Section~\ref{sec:quantum}).

\begin{figure}[t]
\begin{center}
\includegraphics[width=8.6cm]{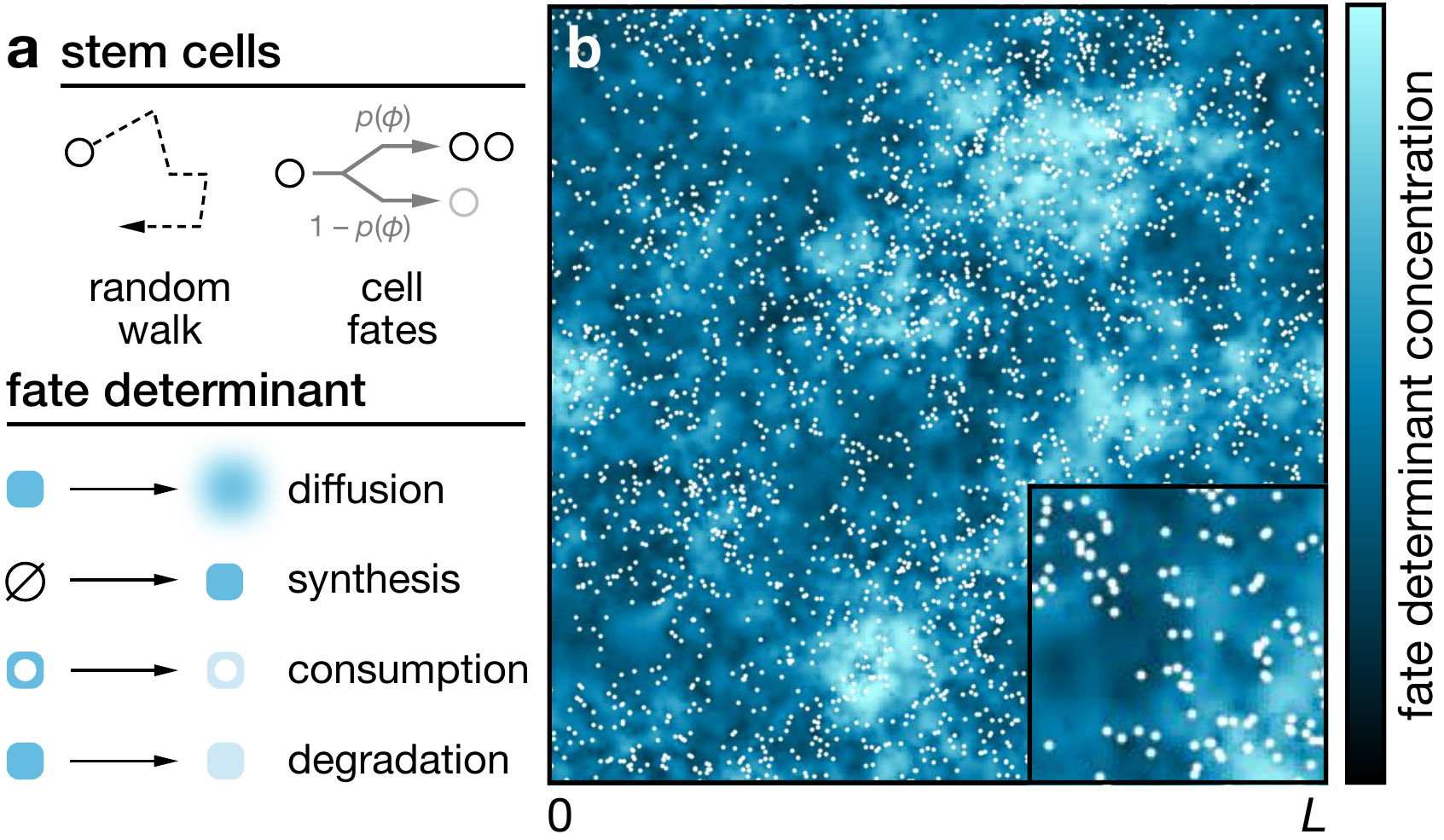}
\caption{
(a) Summary of the individual-based model dynamics specified by Eqs.~(\ref{eq.random.walk}--\ref{eq.mitogen.dynamics}).
(b) Simulation snapshot of the individual-based model. Cells are indicated by dots, the fate determinant concentration is shown as a density plot. Initial and boundary conditions and parameters as in Fig.~\ref{fig.phase.diagrams}a'.
}
\label{fig.showcase}
\end{center}
\end{figure}

\section{Model}
\noindent%
To model cell dynamics, we consider an ensemble of $N$~mobile stem cells, represented by point particles with positions $\vec{x}_i,\hdots,\vec{x}_{N}$.
Here, for simplicity, we suppose that each stem cell undergoes an independent random walk on a two-dimensional domain.
Generalization to lower or higher dimension follows straightforwardly.
During this process, stem cells consume and thereby deplete locally a fate determinant, whose concentration is represented by a continuous field $\phi(\vec{x},t)$.
Stem cells stochastically duplicate through division or are lost through differentiation with a probability that depends on the local concentration of fate determinant.
The fate determinant therefore provides an indirect readout of, and feedback on, the local stem cell density.
The fate determinant itself is diffusible and supplied homogeneously throughout the entire domain.
In general, a fate determinant may promote duplication or differentiation---here, without loss of generality, we consider a duplication-promoting fate determinants.
A summary of all model processes is provided in Fig.~\ref{fig.showcase}a.

To define the random walk dynamics, the positions $\vec{x}_i$ of the stem cells $i=1,\hdots,N(t)$ follow the kinetics
\begin{align}
	\frac{\mathrm{d}}{\mathrm{d}t}{\vec{x}}_i = \sqrt{2 \eta} \vec{\xi}_i(t) \ ,
	\label{eq.random.walk}
\end{align}
where $\eta$ denotes the cell motility and the components of the vector $\vec{\xi}_i=(\xi_i^1,\xi_i^2)$ are drawn at random from a Gaussian white-noise distribution with $\langle \xi_i^\alpha(t)\rangle=0$ and $\langle \xi_i^\alpha(t) \smash{\xi_j^\beta}(t')\rangle=\delta_{ij}\delta_{\alpha\beta}\delta(t-t')$.
Stem cells may duplicate through cell division or become lost through differentiation according to the Poisson processes
\begin{align}
	\mathrm{S} \xrightarrow{p \lambda} \mathrm{S} + \mathrm{S} \ , \qquad
	\mathrm{S} \xrightarrow{(1-p) \lambda} \varnothing  \ ,
	\label{eq.cell.fates}
\end{align}
where $\lambda$ is the limiting rate of stem cell duplication and differentiation. Each duplication process is considered to give rise to two stem cells at the same position.
For a given stem cell $i$, the probability $p$ depends on the local fate determinant concentration $\phi$. For the nonlinear feedback related to fate determinant consumption and processing, we choose a sigmoidal function of the Hill-type \cite{KeenerSneyd2008}, so that the probability $p_i(t)$ that cell~$i$ undergoes duplication is given by
\begin{align}
	p_i(t) = h\bigg(\frac{\phi(\vec{x}_i,t)}{\phi_0}\bigg) \ , \qquad h(x) = \frac{x^n}{1+x^n} \ . \label{eq.hill}
\end{align}
Here the exponent $n$ determines the steepness of the concentration dependence;
later, we will see that the phenomenology of the model is largely insensitive to the specific choice of $n$.
Here, $\phi=\phi_0$ corresponds to the threshold concentration for which $p=1/2$, resulting in (\ref{eq.cell.fates}) taking the form of a critical birth-death process at which the stem cell duplication and differentiation are perfectly balanced.
 
The  concentration field $\phi(\mathbf{x},t)$ of the fate determinant evolves according to the kinetic equation,
\begin{align}
	\frac{\partial\phi}{\partial t} &= D \vec{\nabla}^2 \phi+ \nu J(\vec{x}) - \gamma Q(\phi,\rho)- \kappa \phi  \ , 
	\label{eq.mitogen.dynamics}
\end{align}
where $D$ denotes the diffusion constant of the fate determinant, $\nu$ is its production rate, the function $J$ defines the spatial distribution of its source, $\gamma$ is the limiting consumption rate, the function $Q$ defines the consumption dynamics, $\kappa$ is the decay rate and $\rho(\vec{x},t) = \smash{\sum_{i=1}^{N(t)}} \delta(\vec{x}-\vec{x}_i(t))$ represents the stem cell density.
Unless stated otherwise, we here consider a spatially uniform source, $J(\vec{x})=1$. The consumption function~$Q$ is defined by
\begin{align}
	Q(\phi,\rho)= h\bigg(\frac{\phi}{\phi_0}\bigg) \rho \ ,
\end{align}
where $h$ denotes the Hill function given in Eq.~(\ref{eq.hill}). Here, the appearance of the Hill function in the definition acknowledges the fact that the consumption of fate determinants through endocytosis is affected by saturating receptor binding \cite{Kitadate2019}.

\begin{figure}[t]
\begin{center}
\includegraphics[width=8.3cm]{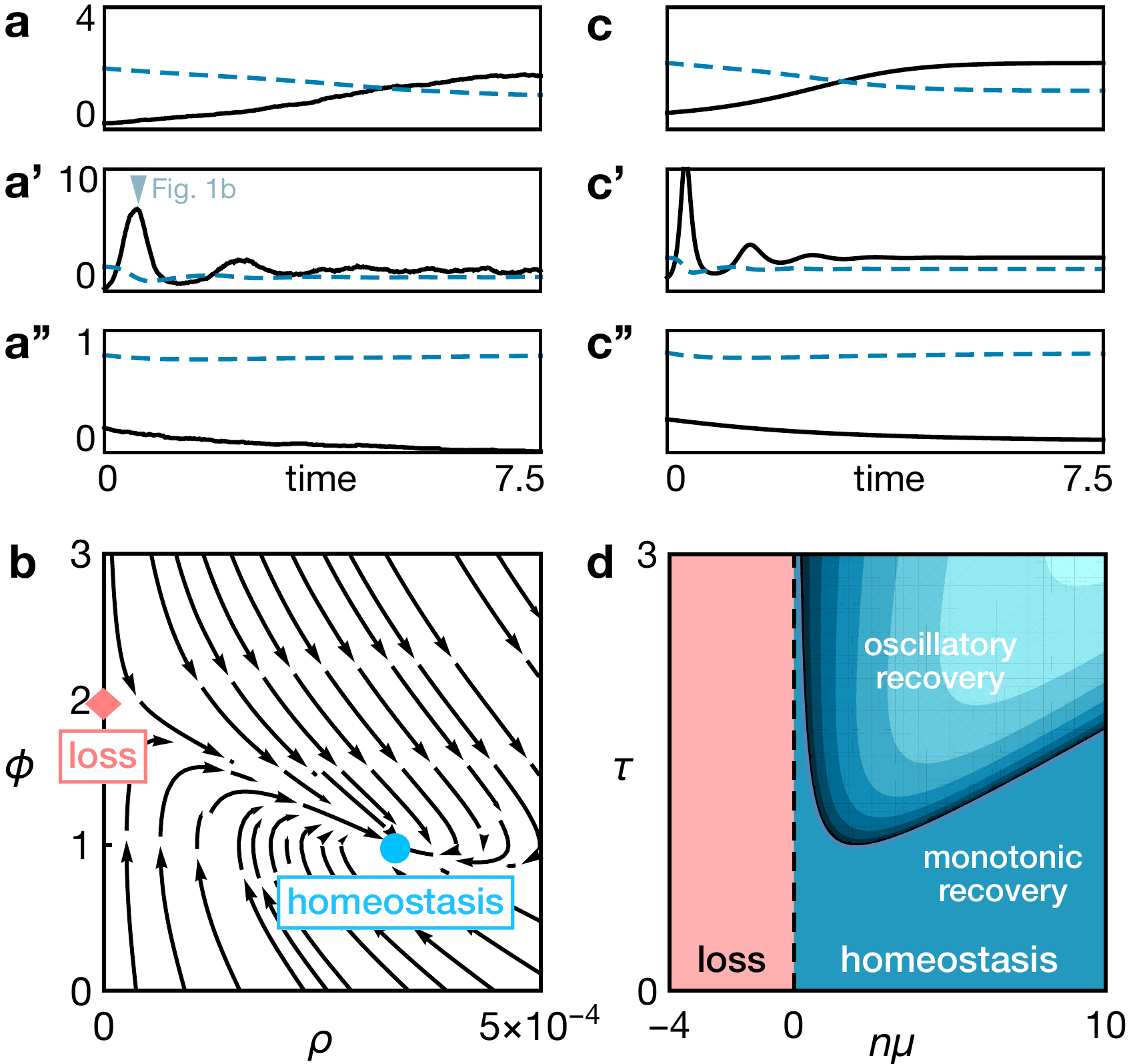}
\caption{
Steady states and transient behavior in the presence of competition for fate determinants.
(a--a') Time evolution of spatial averages of the stem cell density (solid black; rescaled by a factor of $2\times 10^3$) and the fate determinant concentration (dashed blue) from numerical realizations of the model Eqs.~(\ref{eq.random.walk}--\ref{eq.mitogen.dynamics}) for different parameter sets. Parameters are $\eta=25$, $\gamma=10$, $n=2$, $L^2=5000$ and (a) $\lambda=1$, $\nu=2$; (a') $\lambda=20$, $\nu=2$, (a') $\lambda=1$, $\nu=0.8$. Initial conditions: 100 randomly distributed cells, $\phi(\vec{x},0)=\nu$.
The domain has periodic boundary conditions.
(b) Phase portrait of the mean-field model: reaction field $\vec{\Gamma}$, Eq.~(\ref{eq.reaction.field}), homeostatic state $\vec{q}_*$ (blue dot), loss state $\vec{q}_\times$ (red diamond). Parameters as in panel (a).
(c--c') Time evolution of $\rho$ (black solid) and $\phi$ (blue dashed) in the mean-field model. Parameters and conventions as in a--a', respectively.
(d) Phase diagram of the mean-field model, parametrized by the bifurcation parameter $\mu$, Eq.~(\ref{eq.bifurcation.param}), the feedback nonlinearity $n$ as defined in Eq.~(\ref{eq.hill}) and inertial parameter~$\tau$, Eq.~(\ref{eq.inertia}). Colored regions in the oscillatory regime indicate frequencies (from dark to bright in the direction of increasing frequency), see Eq.~(\ref{eq.mf.decay.frequency}).
}
\label{fig.phase.diagrams}
\end{center}
\end{figure}

Fig.~\ref{fig.showcase}b shows a snapshot of a numerical simulation of the system with periodic boundary conditions on a quadratic domain with side length $L$.
Fig.~\ref{fig.phase.diagrams}a--a' shows the time evolution of the spatial averages of the cell density and the fate determinant concentration for different parameter sets (see also Supplementary Movies 1--3).
For all numerical examples, we indicate lengths in units of the fate determinant diffusion length $\sqrt{D/\kappa}$, times in units of the decay time $\kappa^{-1}$ and concentrations in units of the threshold concentration $\phi_0$.
In the appropriate parameter regimes (see below), the system converges to a robust homeostatic state with a well-defined average stem cell density and fate determinant concentration.

\section{Mechanism of homeostasis}
\label{sec:homeostasis}
\noindent%
To gain insight into the phase behavior and dynamics of the system (Fig.~\ref{fig.phase.diagrams}a--a'), we first consider when the homeostatic state is stable, and how its properties such as the homeostatic density depend on the system parameters, including the kinetics of the fate determinant.
To this end, we study a mean-field description of the spatially averaged cell density and fate determinant concentration, which is amenable to an analytical treatment \cite{Kitadate2019}.
This amounts to approximating the stem cell fate dynamics~(\ref{eq.cell.fates}) by a deterministic rate equation for the average density, $\dot \rho = \lambda (2p-1)  \rho$ and taking the spatially averaged mean-field limit of Eq.~(\ref{eq.mitogen.dynamics}).
Defining the density coordinate $\vec{q}=(\rho,\phi)$, the population dynamics is given by
\begin{align}	
	\frac{\mathrm{d}\vec{q}}{\mathrm{d}t}= \vec{\Gamma}(\vec{q}) &\equiv \left( \begin{array}{c}
	 \lambda(2h(\phi/\phi_0)-1)  \rho \\ \nu - \kappa \phi - \gamma h(\phi/\phi_0) \rho
	\end{array} \right)  \ ,
	\label{eq.reaction.field}
\end{align}
with a phase portrait shown in Fig.~\ref{fig.phase.diagrams}b.
This system has two equilibria satisfying the relation $\vec{\Gamma}(\vec{q})=0$: a homeostatic steady state
\begin{align}
	\vec{q}_* = \left(\!
	\begin{array}{c} \rho_* \\ \phi_*
	\end{array}\!\right) = \left(\!
	\begin{array}{c} 2(\nu-\kappa \phi_0)/\gamma \\ \phi_0
	\end{array}\!\right) \ ,
	\label{eq.homeostatic.state}
\end{align}
and a `loss' state
\begin{align}
	\vec{q}_\times= \left(\!
	\begin{array}{c} 0\\ \nu/\kappa
	\end{array}\!\right) \ . \label{eq.loss.state}
\end{align}
The homeostatic cell density $\rho_*$ is correlated linearly with the supply rate of the fate determinant, $\nu$, a hallmark behavior recently observed in the context of murine spermatogenesis \cite{Kitadate2019}.
Interestingly, the homeostatic concentration $\phi_*$ of the fate determinant does not depend on any of its rate parameters---it is simply defined by the threshold concentration $\phi_0$ for which differentiation and loss of stem cells is perfectly balanced by duplication ($p=h(1)=1/2$), a result imposed by the stationarity of the stem cell density.

The homeostatic state $\vec{q}_*$ is stable as long as the concentration $\phi_*$ is large enough to instruct the corresponding density $\rho_*$ of stem cells. Defining the critical production rate $\nu_\mathrm{c}=\kappa\phi_0$ and the dimensionless parameter
\begin{align}
	\mu = \frac{\nu-\nu_\mathrm{c}}{\nu_\mathrm{c}} \ ,
	\label{eq.bifurcation.param}
\end{align}
the state $\vec{q}_*$ is linearly stable for $\mu>0$ and unstable for $\mu<0$, and vice versa for the loss state $\vec{q}_\times$ (see Appendix~\ref{appendix:stability.analysis} and Fig.~\ref{fig.phase.diagrams}d). At $\mu=0$, both states exchange their stability via a transcritical bifurcation. 
Notably, the bifurcation parameter depends on the kinetic and threshold parameters of the fate determinant alone, emphasizing its importance in maintaining homeostasis.
In the stable region ($\mu>0$), perturbations from the homeostatic state either decay monotonically or with damped oscillations.
Fig.~\ref{fig.phase.diagrams}d shows the phase diagram for the mean-field dynamics, indicating the stability of the homeostatic and loss states, and the domain over which the approach to steady-state is characterized by damped oscillations.

Whether such transient oscillations occur depends on the inertia of the feedback mechanism, as indicated by the ratio of the limiting stem cell cycle rate and the decay rate of the fate determinant,
\begin{align}
	\tau=\frac{\lambda}{\kappa} \ .
	\label{eq.inertia}
\end{align}
Damped oscillations occur when this inertia exceeds a certain threshold,
\begin{align}
	\tau > \frac{(n\mu+2)^2}{8n\mu} \ , \label{eq.mf.oscillations}
\end{align}
i.e., when the half-life of the fate determinant is long as compared to the limiting duplication/loss rate of the stem cells.
In the linearized regime, the (angular) frequency of the oscillations and their exponential decay time $T_*$ are given by
\begin{align}
	\omega = \sqrt{\frac{8 n \mu \lambda}{\kappa} - (n\mu+2)^2 }\frac{\kappa}{4} \ , \quad
	T_* = \frac{4}{(n\mu+2)\kappa} \ ,
	\label{eq.mf.decay.frequency}
\end{align}
(see Appendix~\ref{appendix:stability.analysis}). The frequency $\omega$ increases with the limiting cell cycle rate $\lambda$ but has a nonmonotonic dependence on the supply with fate determinants, encoded by the bifurcation parameter $\mu$; notably, the time scale $T$ of oscillation decay has no contribution from the limiting cell cycle rate $\lambda$ close to the homeostatic state.
If perturbed outside the linearized regime, the system can give rise to transient huge-amplitude oscillations (Fig.~\ref{fig.phase.diagrams}a',c' and Supplementary Movie 2). 
Persistent oscillations, i.e., strictly periodic solutions, do not exist in the spatial mean-field description (see Appendix~\ref{appendix:no.limit.cycles}).

In an experimental context, such transient oscillations during a recovery phase provide a hallmark of the competition mechanism and a means to locate the model in parameter space.
Notably, such pronounced oscillations during the recovery from artificially induced injury scenarios have been observed experimentally in the context of murine spermatogenesis \cite{Kitadate2019}.
Similarly, mouse hematopoietic stem cells are known to show a massive overshoot above the steady state after injury through cytotoxic drugs \cite{DeWys1970}.

\begin{figure}[t]
\begin{center}
\includegraphics[width=8.6cm]{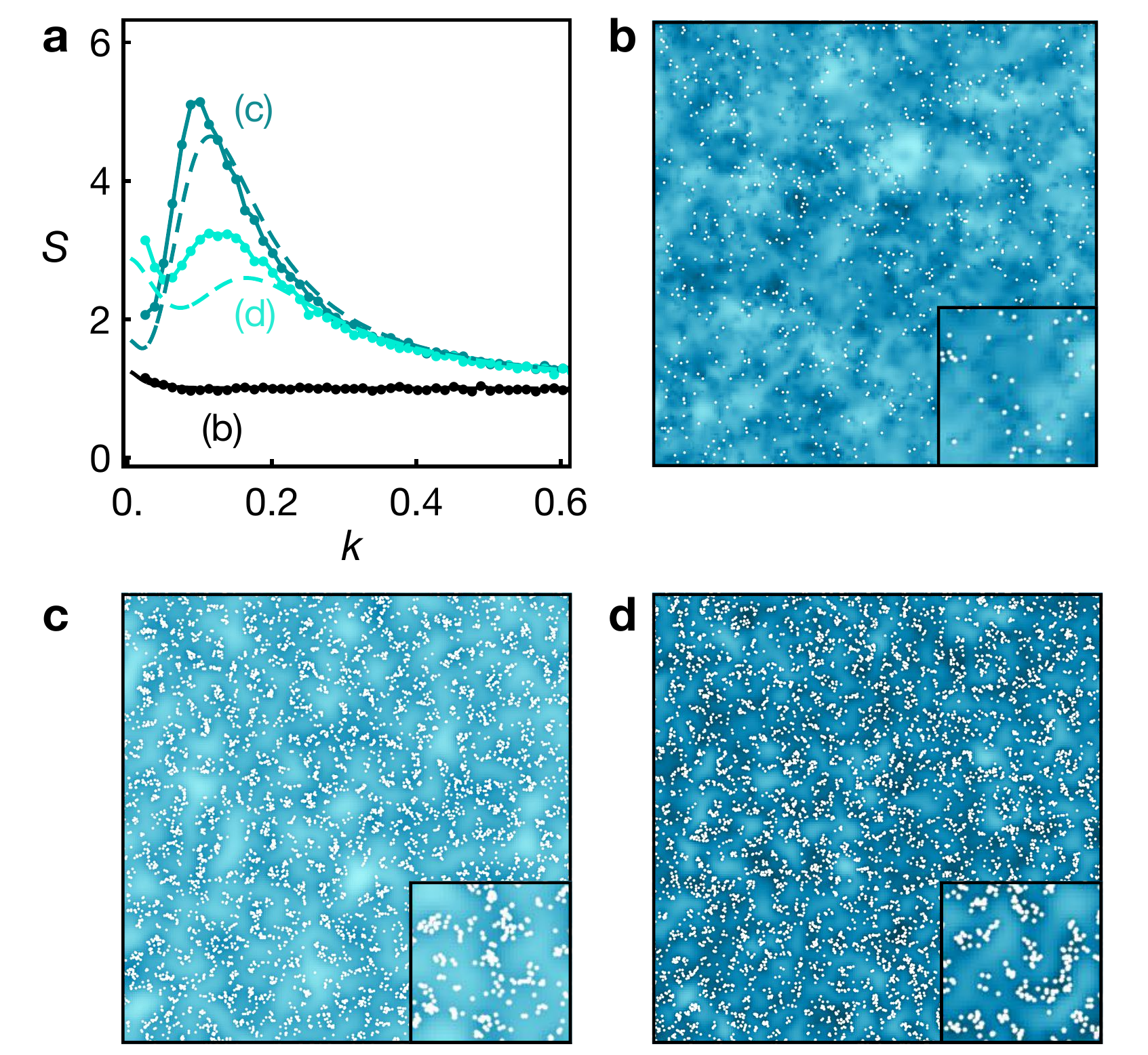}
\caption{
Characterization of non-equilibrium density fluctuations.
(a) Structure factor $S(k)$, given by Eq.~(\ref{eq.sf.def}) averaged over time and wavevectors with $|\mathbf{k}|=k$, for three different systems, for which snapshots are shown in panels b--d. Dashed curves show the analytical approximation Eq.~(\ref{eq.sf}). Parameters are the same as that used in Fig.~\ref{fig.phase.diagrams}a except for
(c,d) $\nu=17$, $\gamma=20$ with
(c) $\eta=0.01$, $\lambda=40$, $\gamma=1$;
(d) $\eta=0.1$, $\lambda=80$, $\gamma=0.2$.
According to the mean-field criterion~(\ref{eq.mf.oscillations}), system (b) is in the monotonic recovery regime, whereas systems (c,d) are in the oscillatory recovery regime.
Diffusion and degradation rates of the fate determinant have been adjusted such that the domains show the same absolute size.}
\label{fig.sf}
\end{center}
\end{figure}

\section{Stem cell density fluctuations}
\label{sec:fluctuations}
\noindent%
In homeostatic tissues, the stem cell density shows a striking degree of large-scale homogeneity, despite non-trivial local density fluctuations. Indeed, in studies of mouse spermatogenesis, stem cells were shown to be characterized by a remarkably homogeneous distribution throughout the seminiferous tubules, as well as clustering-like local fluctuations on small length scales \cite{Kitadate2019}.
Likewise, simulations of our system display a largely constant cell density on long length scales (Fig.~\ref{fig.sf}).
This large-scale homogeneity is in stark contrast with the behavior of systems composed of pure random walkers with an intrinsically balanced self-renewal program ($p=1/2$) without local feedback, i.e., an `engrained' critical birth-death process.
In fact, in such systems, all density correlations have been shown to diverge, leading to a phenomenon termed `neutral clustering' \cite{Houchmandzadeh2002,Houchmandzadeh2008,Houchmandzadeh2009}, which would translate to massive local agglomerations separated by vast regions devoid of stem cells. This situation would be undesirable if tissue integrity is to be maintained, and accordingly, such divergences must be prevented by an adequate homeostatic feedback mechanism such as that presented by the current model.

Since systems of interacting individuals out of equilibrium are known to exhibit characteristic hallmarks of their underlying dynamics in their density fluctuations \cite{Ramaswamy2003,Narayan2007,Yamaguchi2017,Hannezo2017}, it is instructive to study the corresponding fluctuation spectra of the competition model. Further, we will contrast it to alternative models such as the aforementioned scheme which assumes a cell-intrinsic stochastic fate balance without feedback \cite{Klein2011}.

Here we assess the density fluctuation spectrum of the model in the vicinity of the homogeneous homeostatic state through analysis of the static structure factor
\begin{align}
	S(\mathbf{k}) = \bigg\langle \frac{1}{N}  \bigg|\sum_{i=1}^N\mathrm{e}^{\mathrm{i} \mathbf{k}\cdot\mathbf{x}_i}\bigg|^2 \bigg\rangle \ ,
	\label{eq.sf.def}
\end{align}
where $\mathbf{k}$ is the wavevector and the $\mathbf{x}_i$ with $i=1,\hdots,N$ denote the cell coordinates and the average runs over realizations of the system at steady-state.

\begin{figure}[t]
\begin{center}
\includegraphics[width=8.6cm]{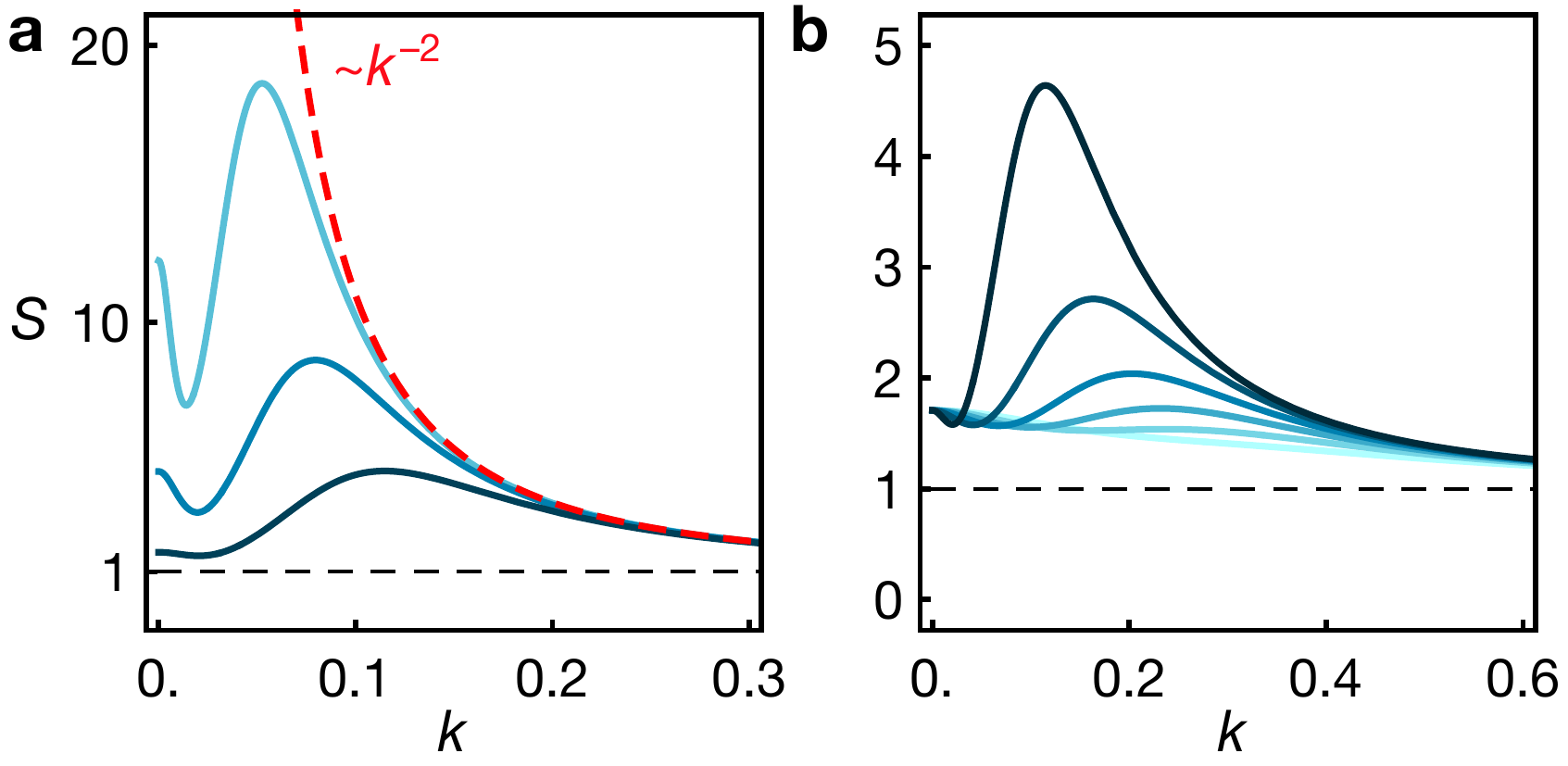}
\caption{
Dependence of the fluctuation spectrum on the feedback strength and diffusion of the fate determinant.
(a) Structure factor $S(k) = S(|\vec{k}|)$, Eq.~(\ref{eq.sf}), for different values of the exponent $n$ indicating the feedback nonlinearity: $n=0$ (red dashed), corresponding to a critical birth-death process, and $n= 0.1$, $0.5$, $2$ (blue; from bright to dark).
(b) Structure factor for different fate-determinant diffusivities: $D=0.01$, $0.025$, $0.05$, $0.1$, $0.25$, $1$ in multiples of the diffusion constant in Fig.~\ref{fig.sf}c (from bright to dark).
All other parameters as in Fig.~\ref{fig.sf}c.
}
\label{fig.sf.peaks}
\end{center}
\end{figure}

Starting with the most basic scenario of pure random walkers without birth or death ($\lambda=0$), i.e., with strictly conserved cell number, the steady-state density distribution is entirely random, resulting in $S(\mathbf{k})=1$ (dashed black line in Fig.~\ref{fig.sf.peaks}).

If we further impose a critical birth-death process that acts intrinsically without any local feedback (i.e., $\lambda\neq 0$ and $n=0$, so that $p \equiv 1/2$), the structure factor is given by $S(\mathbf{k}) = S_0(\mathbf{k})$ with
\begin{align}
	S_0(\mathbf{k})
	&=  1 + \frac{\lambda}{2\eta \mathbf{k}^2} \ ,
	\label{sf.cbd}
\end{align}
which entails a divergence $S_0(\mathbf{k}\to 0) \propto \mathbf{k}^{-2}$ at large length scales, signaling the emergence of giant density fluctuations \cite{Houchmandzadeh2002} (dashed red curve in Fig.~\ref{fig.sf.peaks}a). (For details, see Appendix~\ref{appendix:structure.factor}.)

Simulations show that the proposed competition mechanism is located between these two extreme scenarios---it prevents neutral clustering, but does give rise to non-divergent clustering at a characteristic length scale depending on the system parameters, as evidenced by peaks of the structure factor at finite length scales (Figs.~\ref{fig.sf} and \ref{fig.sf.peaks}).
To provide analytical insight into the mode structure of the density fluctuations, we obtain a density approximation for the stem cell dynamics, which yields the stochastic field theory,
\begin{align}
\begin{split}
	\frac{\partial\rho}{\partial t} &= \eta \vec{\nabla}^2 \rho + \lambda[2h(\phi/\phi_0)-1] \rho \\
	&\qquad + \vec{\nabla} \cdot  \sqrt{2\eta\rho}\vec{\xi} + \sqrt{\lambda \rho} \zeta \ ,
\end{split} \label{eq.sct1} \\[4pt]
	\frac{\partial \phi}{\partial t} &= D \vec{\nabla}^2 \phi + \nu - \gamma {h}(\phi/\phi_0) \rho - \kappa \phi \ , \label{eq.sct2}
\end{align}
where $\zeta$ and the components $\xi_i$ of the vector $\vec{\xi}=(\xi_1,\xi_2)^\mathrm{T}$ are drawn at random from mutually uncorrelated Gaussian white-noise distributions with unit variance (see Appendix~\ref{appendix:structure.factor}).
We first note that the deterministic part of the field theory, Eqs.~(\ref{eq.sct1}) and (\ref{eq.sct2}), does not exhibit any patterning instabilities in the homeostatic regime: For $\mu>0$, the homogeneous state $(\rho(\vec{x},t),\phi(\vec{x},t))=(\rho_*,\phi_*)$, given by Eq.~(\ref{eq.homeostatic.state}), is always stable, as shown in Appendix~\ref{appendix:patterning.instabilities}.
Using Eqs.~(\ref{eq.sct1}) and (\ref{eq.sct2}), we compute an approximation for the static structure factor in the vicinity of the homeostatic state, which is given by
\begin{align}
\begin{split}
	S(\mathbf{k}) &= \frac{(2u_\vec{k}+\lambda)(\omega^+_\vec{k}-\omega^-_\vec{k})(2 v_\vec{k}^2+\omega^+_\vec{k}\omega^-_\vec{k})}{\sqrt{32} \omega^0_\vec{k}(u_\vec{k}+v_\vec{k})(u_\vec{k} v_\vec{k}+\Omega^2)}\ ,
\end{split} \label{eq.sf}
\end{align}
where we have defined $\Omega = \sqrt{\lambda \kappa n \mu/2}$ and
\begin{align}
\begin{split}
	u_\vec{k} &= \eta\vec{k}^2 \\
	v_\vec{k} &=D\vec{k}^2+(1+n\mu/2 )\kappa \ , \\
	\omega^0_\vec{k} &= \sqrt{(u_\vec{k}-v_\vec{k})^2-4\Omega^2}\ , \\
	\omega^\pm_\vec{k} &= \sqrt{\smash{u_\vec{k}^2+v_\vec{k}^2}  \pm (u_\vec{k}+v_\vec{k})\smash{\omega^0_\vec{k}}-2\Omega^2} \ , 
\end{split} \label{eq.sf.aux}
\end{align}
(see Appendix~\ref{appendix:structure.factor}).

Eq.~(\ref{eq.sf}) reveals that, on small length scales, the fluctuation spectrum behaves like that of a critical birth-death process with higher order corrections,
\begin{align}
	S(\mathbf{k}) = S_0(\mathbf{k}) + \frac{\Omega^2}{(D \eta +\eta^2)\vec{k}^4} + \mathcal{O}(\vec{k}^{-6}) \ ,
\end{align}
with $S_0(\mathbf{k})$ given by Eq.~(\ref{sf.cbd}).
On long length scales, the presence of the homeostatic feedback ($n \neq 0$) leads to a homogenization of the cell density and therefore prevents giant density fluctuations, signaled by $S(\mathbf{k}\to 0)$ being finite (Fig.~\ref{fig.sf.peaks}a and Appendix~\ref{appendix:structure.factor}).
In addition, depending on the parameter combination, the structure factor $S(\mathbf{k})$ can exhibit a peak at a characteristic wavenumber $|\mathbf{k}|=k_0$, indicating clustering at a characteristic length scale $2\pi/k_0$ (Figs.~\ref{fig.sf}a and \ref{fig.sf.peaks}).

Generically, clustering is promoted by faster diffusion of the fate determinant (Fig.~\ref{fig.sf.peaks}b), which leads to a more mean-field like behavior. This, in turn, prevents a local response to density deviations, reminiscent of the system of random walkers with an intrinsically critical birth-death process.
On the other hand, the structure factor Eq.~(\ref{eq.sf}) does not depend on the limiting consumption rate of the fate determinant, $\gamma$, indicating its independence of the absolute cell density in the linearized high-density regime.

\section{Clonal evolution through neutral competition}
\label{sec:clonal.dynamics}
\noindent%
In previous studies of stem cell fate behavior, much emphasis has been placed on clonal dynamics of lineage-labelled stem cells, i.e., the size distribution of individually labelled stem cell-derived cohorts, or `clones', resolved over time. This property can also be readily accessed experimentally using transgenic animal models \cite{Klein2011,Blanpain2013,Rulands2018}. Such distributions have been shown to encode features of the underlying pattern of stem cell self-renewal through universal statistical scaling behaviors of clone size.
Stochastic duplication and loss leads to a continuous extinction of clones until the system reaches monoclonality (as shown in Supplementary Movie~4, where clones are marked with different colors).

To analyze the time evolution of stem cell clones, we study the behavior of the cumulative clone size distribution $C(s,t)$, i.e., the probability that a surviving clone derived from a single stem cell at time $t=0$ has a size of at least $s$ stem cells after a time $t$.
Typically, stochastic cell fate decisions underlying population asymmetry lead to a scaling behavior of the clone size distribution,
\begin{align}
	C(s,t)=g\bigg(\frac{s}{\llangle s(t) \rrangle}\bigg) \ ,
\end{align}
where $\llangle \cdot \rrangle$ denotes the expectation value of the surviving population and $g$ is a scaling function that depends on the nature of the underlying cell fate processes and the dimensionality of the system \cite{Klein2011}.

\begin{figure}[t]
\begin{center}
\includegraphics[width=8.3cm]{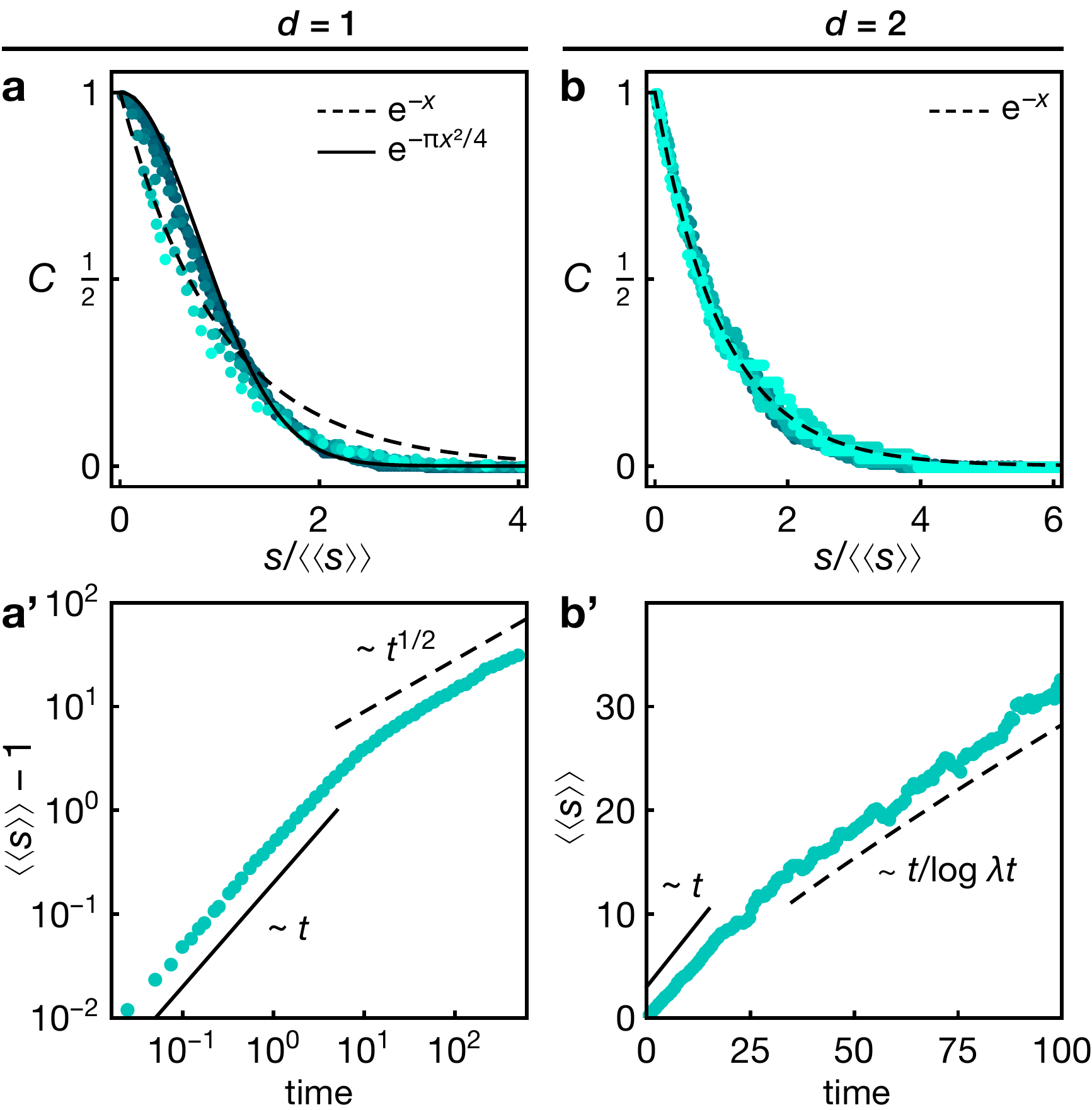}
\caption{
Statistical hallmarks of the clone size distribution.
(a,b) Cumulative surviving clone size distribution $C$ as a function of the scaled clone size from simulations (dots) in $d=1$ and $d=2$ spatial dimensions. Different colors show different time points (increasing time from bright to dark). Black curves show the scaling solutions Eqs.~(\ref{eq.cbd}) and (\ref{eq.vm}) as indicated.
(a',b') Time evolution of the average clone size $\llangle s \rrangle$ of the surviving population for the same systems as shown in panels a and b, respectively. Parameters as in Fig.~\ref{fig.phase.diagrams}a except for $\eta=0.25$, $\gamma=0.25$.
}
\label{fig.clones}
\end{center}
\end{figure}

In systems characterized by a one-dimensional geometry, mechanisms of self-renewal based on local environmental regulation over a characteristic interaction range have been shown to lead to a temporal crossover between scaling regimes~\cite{Yamaguchi2017}:
At early times, when typical clone sizes are small relative to the interaction range, the clone size distribution and time evolution of the average clone size shows the signature of an intrinsically critical birth-death process (i.e., the process (\ref{eq.cell.fates}) with $p=1/2=\mathrm{const.}$) whereas, at later times, when the average clone size has exceeded the interaction range, the system effectively behaves like a `voter model'~\cite{Clifford1973}, in which neighboring cells are stochastically lost and replaced.

We then find, through simulations of the current model, that homeostasis through clonal competition for fate determinants recapitulates this behavior, both in one and two spatial dimensions (Fig.~\ref{fig.sf}).
At early times, the system effectively behaves like a critical birth-death process, which is characterized by the scaling solution
\begin{align}
\begin{split}
	g(x)&=\mathrm{e}^{-x} \ ,\\
	\llangle s(t) \rrangle &= 1+\lambda t/2\ .
\end{split} \label{eq.cbd}
\end{align}
At later times, when the average clone size has exceeded the interaction range, the system behaves like a voter model, whose scaling function $g$ and time dependence of the average surviving clone size $\llangle s(t) \rrangle$ depend on the dimensionality $d$ of the system \cite{Sudbury1976,Sawyer1979,Bramson1980,Klein2010,Klein2011},
\begin{align}
\begin{split}
	g(x) &= \begin{cases}
		\mathrm{e}^{-\pi x^2/4} & d = 1 \\
		\mathrm{e}^{-x}  & d = 2
	\end{cases} \ , \\
	\llangle s(t) \rrangle &= \begin{cases}
		\sqrt{\lambda t} & d = 1 \\
		\lambda t/\ln \lambda t   & d = 2
	\end{cases} \ .
\end{split} \label{eq.vm}
\end{align}
Fig.~\ref{fig.clones} shows rescaled clone size distributions and the average clone size for different time points on a strictly one-dimensional domain (Fig.~\ref{fig.clones}a,a') and in two dimensions (Fig.~\ref{fig.clones}b,b').
In one dimension, not only the average clone size but also the clone size distribution shows a crossover behavior between two different scaling regimes; in two dimensions, the scaling function of the critical birth-death process and the voter model are indistinguishable, cf.~Eqs.~(\ref{eq.cbd}) and (\ref{eq.vm}).
In summary, the competition model is compatible with well-known and experimentally observed clonal dynamics in many tissue types.

\section{Colonization dynamics}
\label{sec:colonization}
\begin{figure}[t]
\begin{center}
\includegraphics[width=8.6cm]{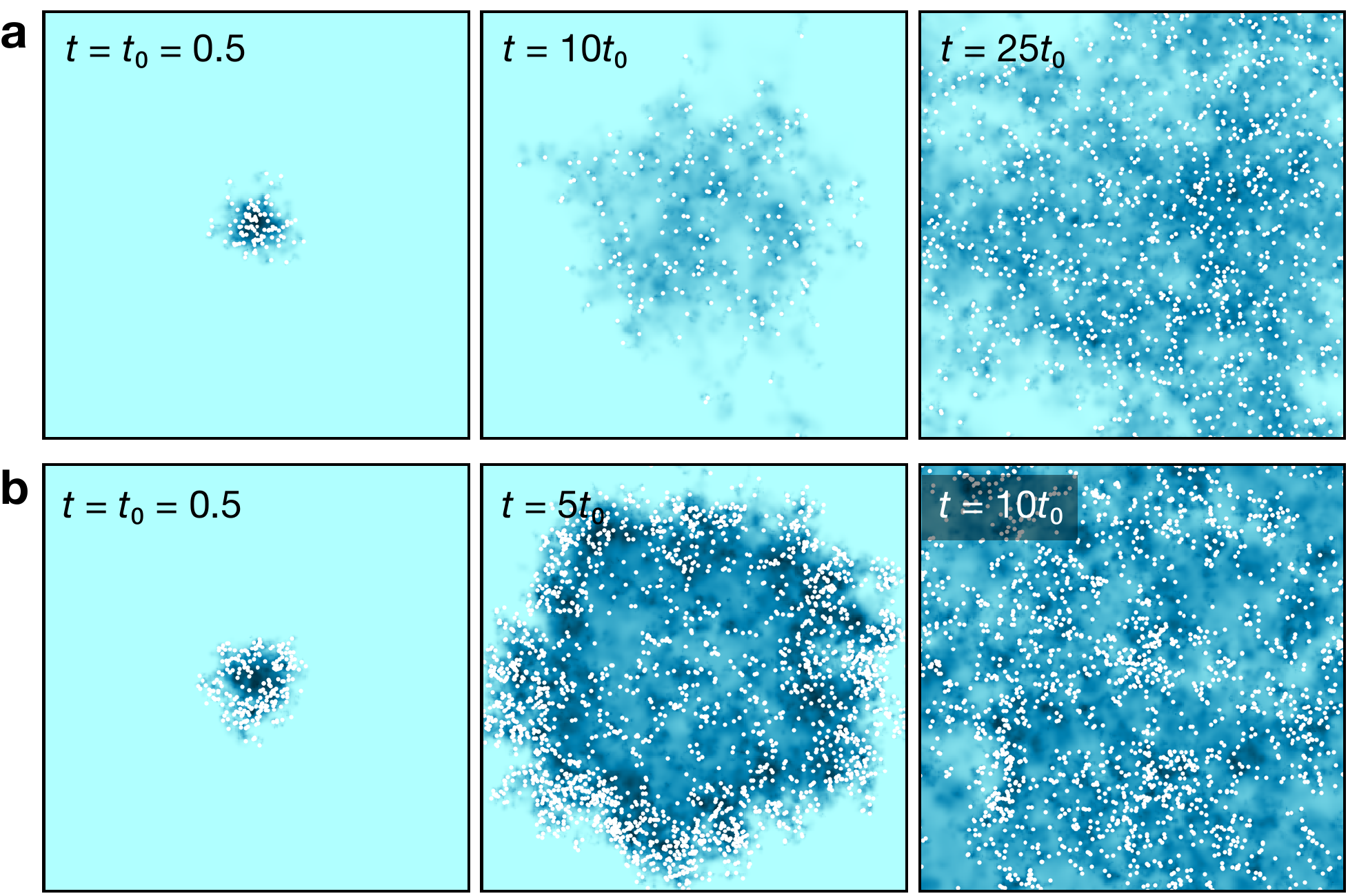}
\caption{
Travelling density waves emerge during colonization scenarios. Plots show a snapshot of the individual-based model at indicated time points.
Cell and fate-determinant density representation as in Fig.~\ref{fig.showcase}a.
Parameters are (a) as in Fig.~\ref{fig.phase.diagrams}a, (b) as in Fig.~\ref{fig.phase.diagrams}a' except for $n=1$ and $L^2=11250$ in both cases. Initial conditions: 100 cells randomly distributed on a disc with radius $0.05 L$ and $\phi(\vec{x},t)=\nu$. The domain has periodic boundary conditions.
}
\label{fig.colonization}
\end{center}
\end{figure}

\noindent%
While clonal competition leads to a voter model-like behavior, free expansion of a clone into a tissue devoid of stem cells, i.e., a \emph{de novo} colonization of a tissue, shows a qualitatively different behavior within the current model paradigm.
Despite individual cells performing independent random walks, the expanding cell population forms a traveling density front that invades the system with a constant mean velocity when averaged over realizations (Fig.~\ref{fig.colonization} and Supplementary Movies 5 and 6).
If the parameters of the system are located within the phase of oscillatory recovery (Fig.~\ref{fig.phase.diagrams}a'), the system exhibits a wave of increased density at the front of the invading population (Fig.~\ref{fig.colonization}b).
In the wake of the invading front, a homeostatic steady state density is left behind.

\begin{figure}[t]
\begin{center}
\includegraphics[width=8.3cm]{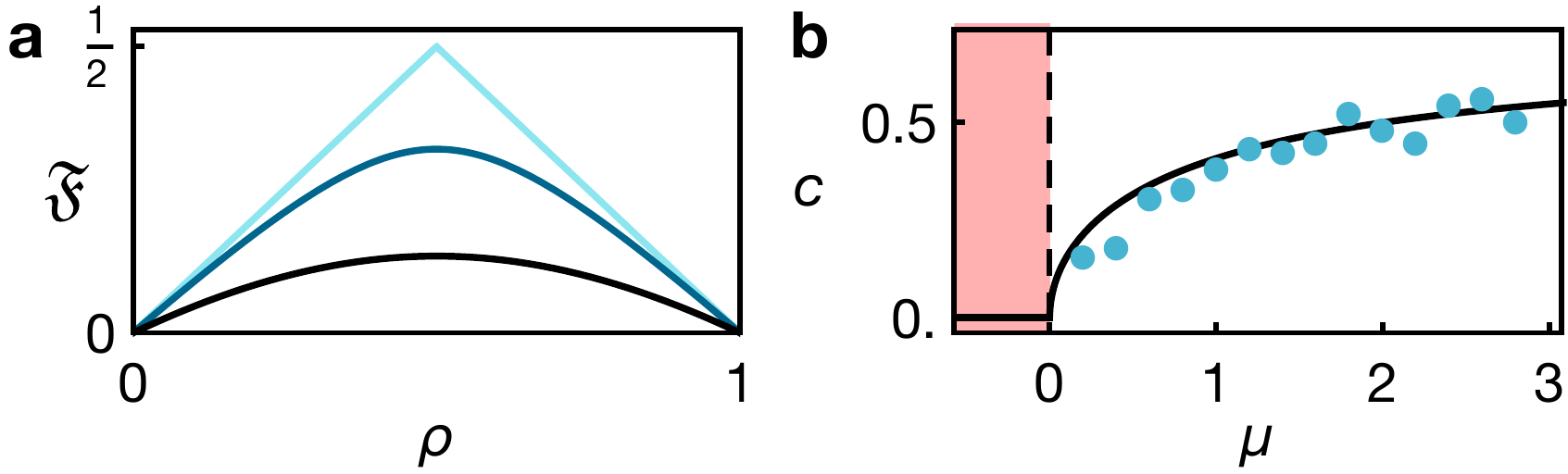}
\caption{
(a) Effective reaction function $\mathfrak{F}$, Eq.~(\ref{eq.fisher.reaction}), mediating the feedback of the competition mechanism on the cell density for $\rho_0=0.5,0.55,1$ (from bright to dark).
(b) Front speed velocity $c$ as a function of the bifurcation parameter $\mu$ from simulations (dots) and the quasi-static approximation Eq.~(\ref{eq.front.speed}). Simulation parameters as in Fig.~\ref{fig.colonization}a. In simulations, $\mu$ has been varied by varying the production rate $\nu$ of the fate determinant.
}
\label{fig.front}
\end{center}
\end{figure}

To gain insight into the nature of the density front propagation, we consider a continuum theory of the system based on a quasi-static approximation of the dynamics of the fate determinant, which yields the effective evolution equation for the stem cell density,
\begin{align}
\begin{split}
	\frac{\partial\rho}{\partial t} &= \vec{\nabla}^2 \rho + \mathfrak{F}(\rho) \ ,
\end{split} \label{eq.fisher}
\end{align}
where the function $\mathfrak{F}$ effectively mediates the competition mechanism (see Appendix~\ref{appendix:quasi.static}). For simplicity, here we choose a feedback exponent $n=1$ for the Hill function Eq.~(\ref{eq.hill}), for which we obtain the reaction function
\begin{align}
\begin{split}
	\mathfrak{F}(\rho) &= \rho_0 - \sqrt{\rho_0^2 - \rho(1-\rho) } \ ,
\end{split} \label{eq.fisher.reaction}
\end{align}
where $\rho_0 = 1/2+\mu^{-1}$ (Fig.~\ref{fig.front}a).
The reaction-diffusion-like system described by Eqs.~(\ref{eq.fisher}, \ref{eq.fisher.reaction}) is of the Fisher--KPP type (see Appendix~\ref{appendix:quasi.static}). Hence, the system is able to exhibit propagating front solutions of the form $\rho(x,t)=\bar\rho(x+ct)+\bar\rho(x-ct)$.
For a localized initial condition, the front speed is given by $c=2 \smash{\sqrt{\mathfrak{F}'(0)}}$ \cite{Cencini2003}, which, restoring dimensions, equates to
\begin{align}
	c = \sqrt{\frac{4\eta\lambda}{1+2\mu^{-1}}} \ . \label{eq.front.speed}
\end{align}
Hence, in the quasi-static limit, as the system approaches the point of phase bifurcation $\mu \to 0^+$, we find a power-law decay of the front speed as $c \propto \mu^{1/2}$; whereas, deep in the active phase ($\mu \gg 0$), the front speed only depends on the cell motility and the division rate, $c \sim (\eta\lambda)^{1/2}$.

Fig.~\ref{fig.front}b shows a comparison of the predicted front speed dependence, Eq.~(\ref{eq.front.speed}), to the average front speed in simulations of the individual-based model far away from the quasi-static limit; the comparison suggests that, even in this case, Eq.~(\ref{eq.front.speed}) provides a viable approximation for the front speed.
Note, however, that this quasi-static approximation cannot account for the density peaks observed in the oscillatory regime (Fig.~\ref{fig.colonization}b), as the limit of infinitely fast dynamics of the fate determinant removes the inertia necessary for the emergence of transient oscillations.

\section{Non-neutral dynamics of \\ mutant stem cell clones}
\label{sec:advantage}
\noindent%
So far, we have considered an equipotent population of stem cells with a homogeneous proliferative potential.
However, in biological tissues, mutations in genes that mediate the sensitivity of the cell fate regulation to fate determinants can alter the proliferative activity or fate potential of the mutated stem cells in a heritable manner \cite{Goriely2003,Goriely2012}.
Such effects can confer a competitive advantage of mutant stem cells over wildtype ones, leading to a progressive colonization of tissue by mutant clones, a process that in general terms is referred to as ``field cancerization'' \cite{Slaughter1953}.
In the present context, experimentally-motivated scenarios in which such mutations are artificially induced can serve as an additional test of the competition mechanism.

To study the impact of the presence of mutant stem cells with a higher competitive potential on the entire tissue, we prepare the system in a homeostatic state for a reference population (termed~`W' for `wildtype').
We then add a localized discrete population of cells with a competitive advantage (termed~`M' for `mutant').
Competitive advantage is formalized through a differential responsiveness to the fate determinant, i.e., individual threshold levels $\phi_0^\mathrm{W}$ and $\phi_0^\mathrm{M}$ with $\phi_0^\mathrm{M}<\phi_0^\mathrm{W}$ entering Eq.~(\ref{eq.hill}) for the respective population.
To reflect the inheritance of mutations, daughter cells of mutant stem cells acquire the same increased susceptibility to the fate determinant. As a result, all stem cells in a mutant clone maintain the same level of competitive advantage over their wildtype neighbors.

\begin{figure}[t]
\begin{center}
\includegraphics[width=8.6cm]{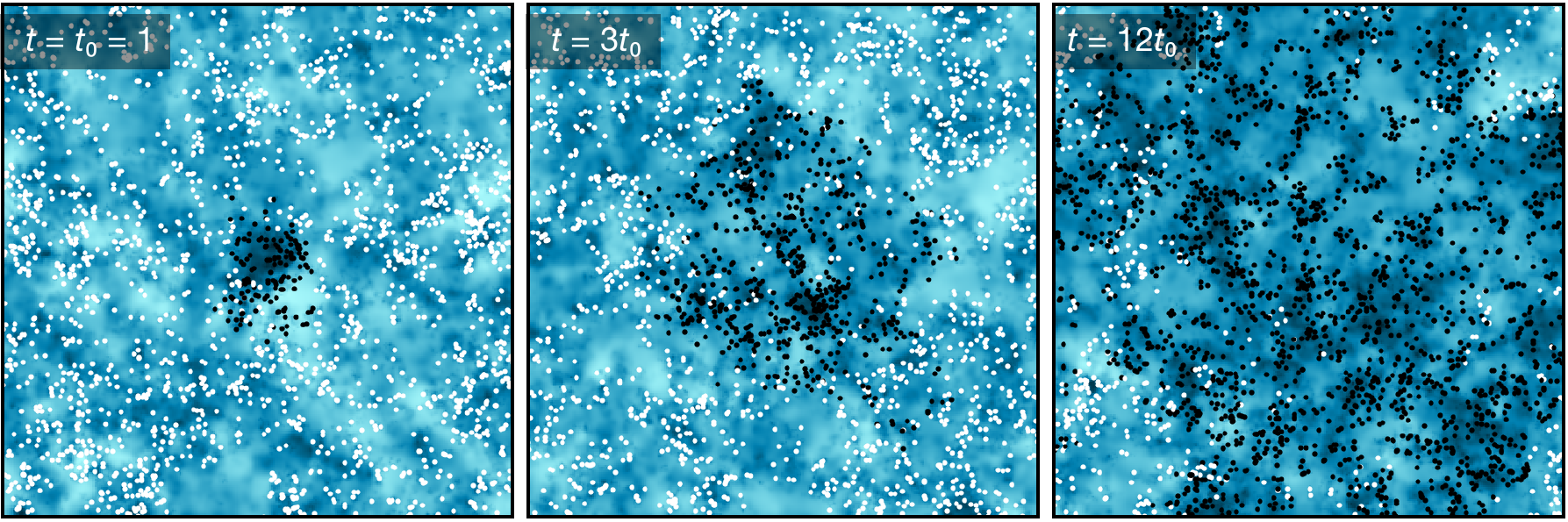}
\caption{
Higher responsiveness of a cell subpopulation~`M' (black) to the fate determinant leads to a takeover of the system and extinction of the reference population~`W' (white).
Plots show a snapshot of the individual-based model at indicated time points; all conventions as in Fig.~\ref{fig.colonization}.
Parameters for the population~`W' as in Fig.~\ref{fig.colonization}b; population~`M' has a threshold concentration $\phi_0^\mathrm{M} = 0.8 \phi_0^\mathrm{W}$. At time $t=0$, 100 mutant stem cells have been inserted in a circular domain of radius $0.05L$ into the steady state of population `W'.
}
\label{fig.advantage}
\end{center}
\end{figure}

Fig.~\ref{fig.advantage} and Supplementary Movie~7 show a numerical example of such a scenario:
even though there is no direct interaction between the two populations (such as proximity-induced cell death or contact inhibition \cite{Klein2011b}), the population~M invades the entire system, leading to the complete extinction of the population~W.
This front-like invasion is reminiscent of the colonization scenario described in Section~\ref{sec:colonization}, in which stem cells expand into a territory devoid of stem cells.
This behavior is due to the fact that the population~M is able to settle towards a higher homeostatic cell density, which lowers the steady-state concentration of the fate determinant to $\phi_0^\mathrm{M}$, leading to an ongoing depletion of population~W through differentiation; a formal argument on a mean-field level is given in Appendix~\ref{appendix:advantage}.

Fig.~\ref{fig.clones.nonneutral} shows the time evolution of the size distribution of surviving clones of the population~M within the defined scenario in $d=1$ and $2$ spatial dimensions.
Notably, size distributions of mutant clones do not acquire a scaling distribution as in the neutral case, cf.~Section~\ref{sec:clonal.dynamics} and Fig.~\ref{fig.clones}.
Numerically, we find that the mean clone size of the surviving population evolves as $\llangle s \rrangle \sim t$ in one dimension and $\llangle s \rrangle \sim t^2$ in two dimensions (Fig.~\ref{fig.clones.nonneutral}b,b') which, considering the front-like expansion of clones, is the expected behavior if the clone radius grows with a constant velocity, so that its area grows like $t^d$;
the associated variance $\llangle s^2 \rrangle-\llangle s \rrangle^2$ grows like $t$ in $d=1$ and like $t^3$ in $d=2$.

\begin{figure}[t]
\begin{center}
\includegraphics[width=8.3cm]{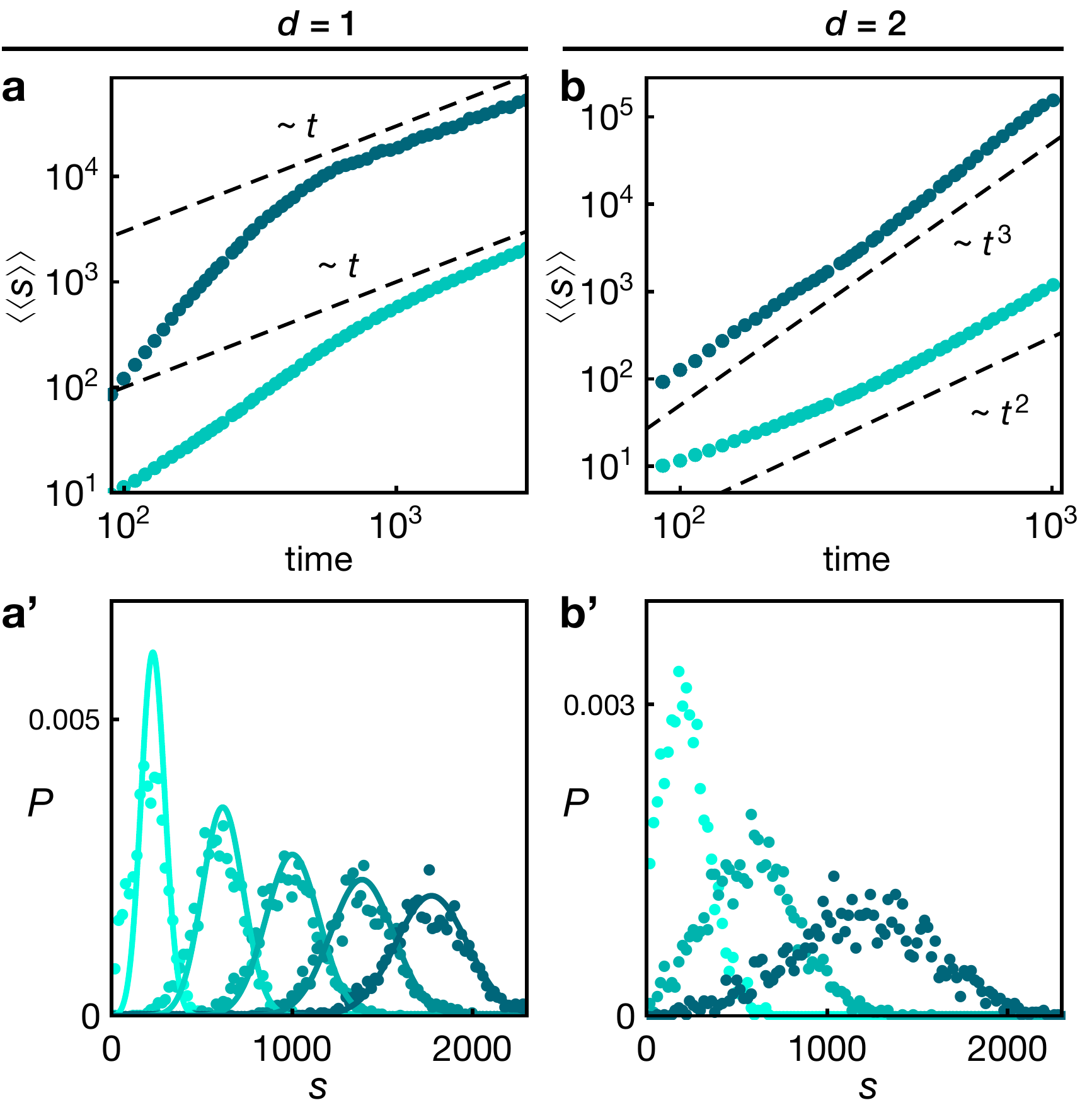}
\caption{
Statistical hallmarks of the clone size distribution for non-neutral competition.
(a,b) Time evolution of the average clone size $\llangle s \rrangle$ (bright) and the variance $\llangle s^2 \rrangle-\llangle s \rrangle^2$ (dark) of the surviving population from simulations (dots) in $d=1$ and $d=2$ spatial dimensions.
(a',b') Surviving clone size distribution $P$ as a function of the clone size for the same systems as shown in panels a and b, respectively. Different colors show different time points (increasing time from bright to dark). Curves show the solutions of the voter model, Eq.~(\ref{eq.vm.expectations}), with parameters determined as described in the main text.
Parameters as in Fig.~\ref{fig.advantage}.
}
\label{fig.clones.nonneutral}
\end{center}
\end{figure}

In Section~\ref{sec:clonal.dynamics}, we have seen that the neutral system exhibits the asymptotic dynamics of a voter model---in the non-neutral case, we therefore expect the asymptotic clonal dynamics to be captured by a \emph{biased} voter model~\cite{Williams1972}, where the bias ${v}$ corresponds to the difference in the loss-and-replacement rates of different populations and is related to the competitive advantage (see Appendix~\ref{appendix:biased.voter.model}).
For $d=1$, the biased voter model can be solved analytically (see Appendix~\ref{appendix:biased.voter.model} and Ref.~\cite{Snippert2013}).
In the asymptotic long-time limit, the probability $P(s,t)$ to find a clone of size $s$ among the surviving population at time~$t$ is given by
\begin{align}
\begin{split}
	P(s,t) &\simeq \sqrt{\frac{ \varepsilon}{2 \pi  r  t}} \frac{1}{1-\mathrm{e}^{-2\varepsilon}} \big(\mathrm{e}^{-\varepsilon (s-1-r t)^2/(2r  t)} \\
	&\hspace{2.6cm} - \mathrm{e}^{-2\varepsilon} \mathrm{e}^{-\varepsilon (s+1-r t)^2/(2r  t)} \big)  \ ,
\end{split} \label{eq.vm.clone.sizes}
\end{align}
where
\begin{align}
	\varepsilon = \ln  \sqrt{\frac{1+{v}}{1-{v}}} \ , \qquad r = 2 \varepsilon \sqrt{1-{v}^2} \ , \label{eq.biased.voter.parameters}
\end{align}
with $v$ being the directional bias and with time being measured in units of the loss-and-replacement time (Appendix~\ref{appendix:biased.voter.model}).
The corresponding mean and variance of the surviving clone size are given by
\begin{align}
\begin{split}
	\llangle s(t) \rrangle &\simeq r t + \frac{1}{\tanh \varepsilon} \ , \\
	\llangle s(t)^2 \rrangle - \llangle s(t) \rrangle^2 &\simeq \frac{r t}{\varepsilon}  - \frac{1}{(\sinh\varepsilon)^2} \ .
\end{split} \label{eq.vm.expectations}
\end{align}
Both quantities evolve linearly in time as is the case for the asymptotic dynamics of the competition model (Fig.~\ref{fig.clones.nonneutral}a).
Hence, by fitting the slopes of the mean and variance, the effective parameters $r$ and $\varepsilon$ can be determined from the individual-based model. Comparing the resulting clone size distributions, Eq.~(\ref{eq.vm.clone.sizes}), of the biased voter model with the numerical results for the individual-based model shows striking agreement (Fig.~\ref{fig.clones.nonneutral}a,a').
In summary, the long-term dynamics of the competition model falls into the voter model class, with competitive advantages being translated as directional biases in the loss-and-replacement process.

\section{A `quantum' tissue}
\label{sec:quantum}

\noindent%
So far, we have only considered systems in which the fate determinant is supplied homogeneously throughout the tissue. However, when this condition is relaxed (as is the case in many biological tissues), interesting effects can emerge. Not only do the stem cells preferentially gather in the vicinity of the sources of the fate determinant (as shown below),
the interplay of random (i.e., diffusive) cellular motion and loss-and-replacement dynamics can lead to a peculiar behavior in the form of a quantization of allowed stationary states.

Formally, this can be seen by realizing that the deterministic part of the field theory for the density dynamics, Eq.~(\ref{eq.sct1}), has the form of an imaginary-time Schr\"odinger equation,
\begin{align}
	\frac{\partial\rho}{\partial t} &= \eta \vec{\nabla}^2 \rho + V(\mathbf{x},t)\rho \ , \label{eq.ischroedinger}
\end{align}
where the kinetic term represents the cell motility and the potential term represents the self-renewal capacity modulated by the presence of the fate determinant; hence, the `potential' $V$ depends on position and time through the concentration field, $V(\mathbf{x},t) = \lambda[2{h}(\phi(\mathbf{x},t)/\phi_0)-1]$.
In contrast to a quantum wavefunction, the density field $\rho$ is real-valued; furthermore, Eq.~(\ref{eq.ischroedinger}) is part of a nonlinear system with the dynamics of $V$ indirectly depending on $\rho$ through the dynamics of $\phi$.
Nevertheless, the dynamics Eq.~(\ref{eq.ischroedinger}) imposes certain quantization conditions enforced by boundary conditions, i.e., through the same mechanism that leads to discrete energy levels in genuine quantum systems.

\begin{figure}[t]
\begin{center}
\includegraphics[width=8.3cm]{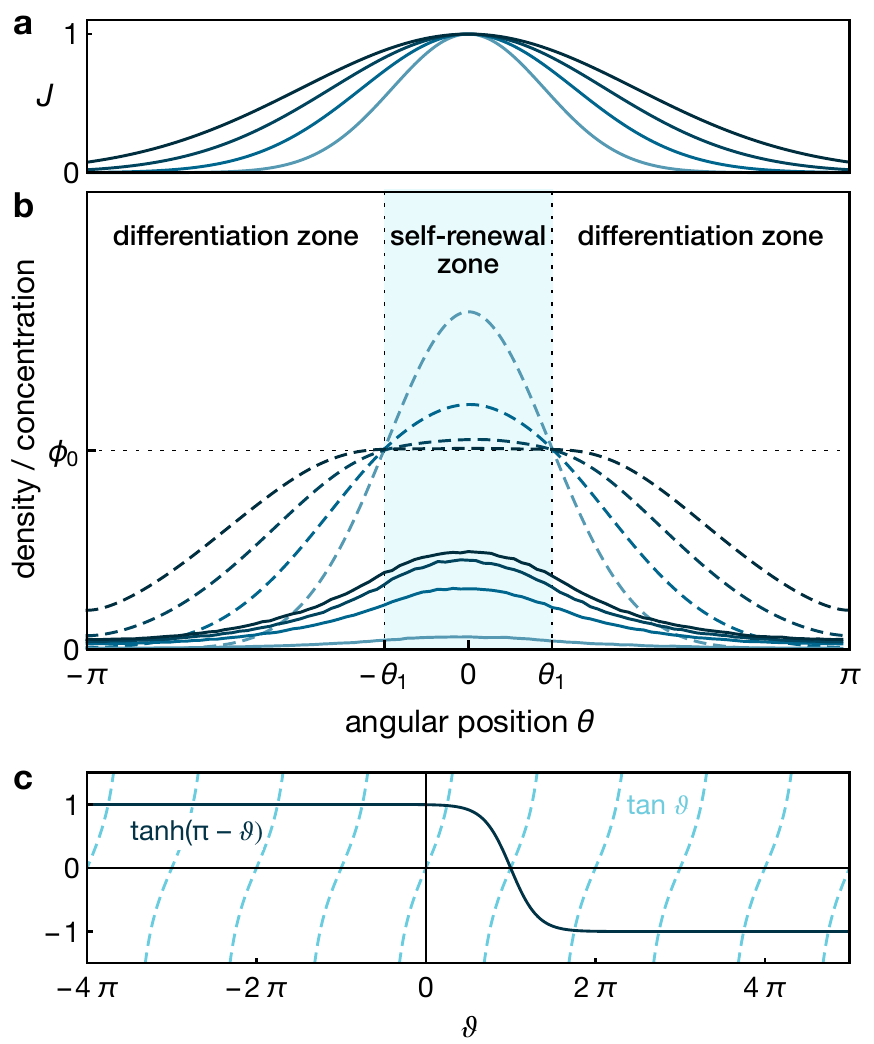}
\caption{
Examples of quantization phenomena on a circular one-dimensional domain.
(a) Spatially extended source regions $J(\theta)$ used in the corresponding simulations shown in panel b. The functional form is a Gaussian $J(\theta)=\smash{\mathrm{e}^{-\theta^2/2\sigma^2}}$ with $\sigma/L=0.10,0.14,0.18,0.22$ (from bright to dark).
(b) Simulation results of the individual-based model for the stationary stem cell density $\rho(\theta)$ (solid; rescaled by a factor of 0.05) and concentration profile $\phi(\theta)$ of the fate determinant (dashed) for the respective source regions shown in panel a.
The blue shaded region indicates where $\phi> \phi_0$.
Vertical dashed lines show the boundary angles $\pm \theta_1$, the smallest solution to Eq.~(\ref{eq.quantization}).
Parameters as in Fig.~\ref{fig.phase.diagrams}a except for $\gamma=10^{-3}$; the domain length is $L=35$.
(c) Graphical representation of the quantization condition: intersections of both curves correspond to solutions to Eq.~(\ref{eq.quantization}).}
\label{fig.quantization}
\end{center}
\end{figure}

A simple but instructive example where such a quantization behavior can be observed---even at the level of the individual-based model---is homeostasis on a one-dimensional circular domain (with angular coordinate $\theta$), in the presence of a localized source $J(\theta)$ of the fate determinant (Fig.~\ref{fig.quantization}a).
Considering the limiting case of a very steep feedback ($n \to \infty$), we find that the stationary state of the individual-based model spatially decomposes into a `self-renewal zone' close to the source, where $\phi(\vec{x}) > \phi_0$, and a `differentiation zone' away from the source, where $\phi(\vec{x}) < \phi_0$ (Fig.~\ref{fig.quantization}b).
The majority of the cells gathers in this self-renewal zone close to the source.
Remarkably, while the overall shape of the stationary solution depends on the width of the source region, the widths of the self-renewal and differentiation zones do not\footnote{In the individual-based model, such a behavior can only be observed if the extension of the cell density distribution around the source is comparable to the system extension, i.e., if the system `feels' the periodic boundary conditions.}.

This idiosyncratic behavior can be traced back to a quantization phenomenon:
Using an analytical approximation (Appendix~\ref{appendix:a.quantum.tissue}), it can be shown that the system can only attain solutions with a certain discrete set of stationary angles $\theta_n$ separating self-renewal and differentiation zones; the $\theta_n$ are proportional to the set of solutions $\vartheta_n$ of the transcendental quantization condition
\begin{align}
	\tan\vartheta=\tanh(\pi-\vartheta) \ , \label{eq.quantization}
\end{align}
(see Fig.~\ref{fig.quantization}c).
Interestingly, the angle $\vartheta$, which separates the self-renewal and differentiation zones, is determined by the dynamics of $\rho$ alone.
While Eq.~(\ref{eq.quantization}) is a necessary condition for the existence of a stationary solution in this one-dimensional circular scenario, it is not sufficient; whether a solution with boundary angle $\theta_n$ exists also depends on the dynamics of the fate determinant.
Since the competition mechanism leads to more plateau-like shapes of the stationary concentration profile of the fate determinant for more extended source regions (Fig.~\ref{fig.quantization}b), the presence of boundary angles with larger $n$ becomes increasingly hard to detect from numerical simulations of the individual-based model.

In summary, when the fate determinant is supplied in localized spatial domains, the competition-based mechanism leads to a preferential localization of the stem cell population to the sources of the fate determinant.
This generically gives rise to structural domains for self-renewal and differentiation, which, under special conditions, can show hallmarks of a quantization behavior.

\section{Discussion}

\noindent%
Here we have introduced a generic model of stem cell density regulation in an open niche environment that is based on competition for a fate determinant that `primes' the stochastic fate behavior of stem cells in a concentration-dependent manner.
Such behavior contrasts models of homeostasis in ecological settings in which competition for nutrients regulates population density by imposing a finite energy supply leading to the starvation of excess populations \cite{Verhulst1845,Schoener1973,Tilman1977,Murray1993,Lejeune2002,Young2009}.
A key phenomenological difference is that in systems governed by competition for fate determinant, the resource that is competed for determines the type but not the amount of progeny generated, as is the case in the stem cell context, where a loss of stem cells occurs through differentiation rather than death.

The competition mechanism enables robust density homeostasis during maintenance and in recovery following severe perturbations from steady-state, such as that imposed by abrupt stem cell depletions and artificial initial conditions created by \emph{de novo} colonization scenarios, followed by regeneration.
We have demonstrated that the long-time clonal dynamics can be captured by the classical `voter dynamics' in one and two dimensions, in accordance with previous studies on the clonal behavior of competing cells \cite{Yamaguchi2017}.
This leads to scaling clone size distributions frequently observed in clonal lineage tracing experiments in various biological tissues \cite{Klein2011}.
Moreover, we have shown that a competitive advantage of a subpopulation of mutant clones entails a departure from the scaling regime, but still gives rise to a power-law time evolution of average quantities, whose exponents depend on the spatial dimensionality.

Finally, we have shown that the competition mechanism can lead to a peculiar quantization of stationary states in the presence of inhomogeneous sources of the fate determinant---a phenomenon that is related to the formal similarity of the stem cell dynamics to the Schr\"odinger dynamics governing quantum wavefunctions, and which is observed even in the individual-based model.
Whether such quantization behavior could be observed \emph{in vivo} depends on the properties of the feedback mechanism, the localization of the fate-determinant sources and the size of bounded domains.

Key features of the model dynamics, such as a linear dependence of the homeostatic stem cell density on the supply of fate determinants, and pronounced density oscillations during recovery from injury, have recently been observed during the maintenance and recovery of the stem cell compartments in murine testis \cite{Kitadate2019}.
This biological system also provides an experimental platform to test further model predictions on density fluctuations, transplantation scenarios and non-neutral clonal competition, combining genetic modifications, intravital live-imaging and clonal lineage tracing \cite{Yoshida2007,Nakagawa2010,Hara2014}.

In tissues maintained by stem cells harbored by `closed' (or `definitive') niches of restricted spatial extension, stem cells compete for limited niche access \cite{Kitadate2007,Morrison2008,Johnston2009,Stine2013,Yoshida2018}.
This scenario can be viewed as a special case of the presented mechanism in which fate determinants are only produced and dispersed in localized domains reflecting the spatially restricted niche regions \cite{Kitadate2019}.
In this regard, the presented mechanism may serve as a more general framework that captures the local rules of both facultative and definitive niche regions.

Our model also shares certain features with biological and ecological models and model systems, while, at the same time, displaying clear conceptual and formal differences.
In contrast to classical reproduction models of random walking agents with built-in criticality \cite{Zhang1990,Young2001,Houchmandzadeh2002,Houchmandzadeh2008,Houchmandzadeh2009}, the competition mechanism prevents diverging density fluctuations and leads to a spatial homogenization of the cell density, although spatial fluctuations with characteristic length scales can occur.
Also, different from many one-component models of competing random walkers, which typically include instantaneous and nonlocal interactions \cite{HernandezGarcia2004,Ramos2008,Heinsalu2010,Heinsalu2012,HernandezGarcia2015}, our entirely local competition model does not exhibit any patterning instabilities in the deterministic limit---features which, in any case, depend on the details of the nonlocal interaction \cite{Pigolotti2007,Pigolotti2009}.
On a formal level, our model also has similarities with models of `chemostats', i.e., bioreactors in which microorganisms compete for a constant supply of nutrients; although these systems are usually well-mixed and therefore do not entail finite-range interactions \cite{Monod1949,Herbert1956,Smith1995,Campillo2011,Campillo2015,Wade2016}.
Many models of tumor growth fall within the same class of nutrient-limited expansion with a similar formal structure; however, in this context, tissues are usually compact and contiguous and mechanical interactions play a key role, which alters the density fluctuation spectrum \cite{Lowengrub2009,Chauviere2012,AlSaedi2018}.
Mechanisms of cross regulation between multiple cell populations as well as microbial `cartels' have been shown to enable the their stable coexistence, behaviors that have been modelled on a population level \cite{Kunche2016,Taillefumier2017,Posfai2017,Adler2018,Zhou2018}.
Interestingly, it has recently been suggested that the maintenance of fibroblast and macrophage populations, two abundant cell types in mammalian tissues, may be based on a related competition mechanism: in this framework, each population produces the growth factors that are consumed by the respective other population, thereby mutually stabilizing their entire population \cite{Adler2018,Zhou2018}.
The competition mechanism proposed here can also be regarded as a form of inverse `quorum sensing' \cite{Miller2001,Doganer2016}, whereby cells detect the density of their neighbors through the \emph{lack} (rather than abundance) of the fate determinant for which they compete.

Our results demonstrate how the interplay of stochastic cell fate dynamics and nonlinear fate control enables a rich variety of dynamical and statistical phenomena. It extends existing mean-field and field-theoretic approaches of homeostasis by taking into account an explicit mediator of fate control whose kinetics have dominant effects on the phase diagram, homeostatic densities, fluctuations and transient phenomena. Even though being motivated by the dynamics of stem cells, this approach is not limited to tissue maintenance as investigated here, but can also be applied to other biological and ecological systems in which motile agents compete for fate-controlling substances.

\begin{acknowledgements}
\noindent%
We thank S.~Rulands, \'E.~Hannezo and H.~Yamaguchi for helpful discussions.
This work was supported by a Wellcome Trust Senior Investigator Award (098357) to B.D.S.
D.J.J.~and B.D.S. acknowledge core funding to the Gurdon Institute from the Wellcome Trust (092096) and Cancer Research UK (C6946/A14492).
S.Y. acknowledges funding through the Grant-in-Aid for Scientific Research (KAKENHI) from JSPS (JP16H02507, JP18H05551), and AMED-CREST from AMED (JP18gm1110005).
\end{acknowledgements}

\begin{appendix}

\section{Stability analysis \\ of the mean-field system}
\label{appendix:stability.analysis}

\noindent%
We obtain the phase diagram of the mean-field model given by Eq.~(\ref{eq.reaction.field}) through a standard linear stability analysis.
Using the definition of the dimensionless parameters~$\mu$ and $\tau$, Eqs.~(\ref{eq.bifurcation.param}) and (\ref{eq.inertia}) and the equilibria given by Eqs.~(\ref{eq.homeostatic.state}) and (\ref{eq.loss.state}), can be written as
\begin{align}
	\vec{q}_* &= \left(\!
	\begin{array}{c} 2\kappa\phi_0 \gamma^{-1} \mu \\ \phi_0
	\end{array}\!\right) \ , \quad
	\vec{q}_\times = \left(\!
	\begin{array}{c} 0 \\ (1+\mu)\phi_0
	\end{array}\!\right) \ .
\end{align}
An equilibrium is stable if the real part of both eigenvalues of the Jacobian matrix $\mathbf{J}$ of $\vec{\Gamma}$ at the respective equilibrium is negative.
The eigenvalues  of $\mathbf{J}(\vec{q}_*)$ are given by
\begin{align}
	e^*_\pm = \frac{\kappa}{2}\bigg({-1}-\frac{n\mu}{2}  \pm \sqrt{ \left(1+\frac{n\mu}{2} \right)^2-2 \tau n \mu } \bigg) \ . \label{eq.mf.eigenvalues}
\end{align}
For $\mu<0$, $\operatorname{Re}(e^*_+)>0$, so that the homeostatic state is unstable; for $\mu>0$, $\operatorname{Re}(e^*_\pm)<0$, so that the homeostatic state is stable. In the latter case,
the eigenvalues acquire an imaginary part if the inequality~(\ref{eq.mf.oscillations}) is satisfied (Fig.~\ref{fig.phase.diagrams}d), and the imaginary part of the eigenvalues~(\ref{eq.mf.eigenvalues}) is given by Eq.~(\ref{eq.mf.decay.frequency}).
The eigenvalues of $\mathbf{J}(\vec{q}_\times)$ are given by
\begin{align}
	e^\times_1 = \left( 1-\frac{2}{1+(1+\mu)^n} \right)\lambda \ , \qquad
	e^\times_2 = -\kappa \ .
\end{align}
Both eigenvalues are always real and, since $\kappa>0$, stability is determined by $e^\times_1$.
For $\mu<0$, $e^\times_1<0$ and the loss state is stable; for $\mu>0$, $e^\times_1>0$ and the loss state is unstable.

\section{Non-existence of limit cycles \\ in the mean-field system}
\label{appendix:no.limit.cycles}

\noindent%
The existence of limit cycles in the mean-field system Eq.~(\ref{eq.reaction.field}) can be ruled out by the Dulac--Bendixson criterion \cite{Burton2005}:
The system has no periodic solution lying entirely within a simply connected region $\mathds{D}$ if there exists a $C^1$ function $\Phi(\vec{q})$, such that the sign of the function $U(\vec{q}) = \vec{\nabla} \cdot \Phi(\vec{q}) \vec{\Gamma}(\vec{q})$, where $\vec{\Gamma}$ is the reaction field defined by Eq.~(\ref{eq.reaction.field}), is constant everywhere within $\mathds{D}$.
Let $\mathds{D}$ be the region $\rho>0$, $\phi \in \mathds{R}$. For the choice $\Phi(\vec{q})=\rho^{-1}$, we obtain $U(\vec{q}) = - \kappa/\rho - (\gamma/\phi_0) {h}'(\phi/\phi_0)$ which, for ${h}'>0$, is negative everywhere within $\mathds{D}$.

\section{Approximation of the \\ static structure factor}

\label{appendix:structure.factor}
\noindent%
Here we compute the approximation Eq.~(\ref{eq.sf}) of the static structure factor $S(\mathbf{k})$ from a stochastic continuum theory approximating the individual-based model given by Eqs.~(\ref{eq.random.walk}--\ref{eq.mitogen.dynamics}).
This continuum theory for the cell density $\rho(\mathbf{x},t)=\sum_i \delta(\mathbf{x}-\mathbf{x}_i)$ and the concentration field $\phi(\mathbf{x},t)$ of the fate determinant is derived by combining an exact density representation of the random walk dynamics Eq.~(\ref{eq.random.walk}) with a system-size expansion of the birth-death process (\ref{eq.cell.fates}) \cite{Dean1999,vanKampen2007,Yamaguchi2017}.
Together with Eq.~(\ref{eq.mitogen.dynamics}) for the dynamics of the fate determinant, this yields Eqs.~(\ref{eq.sct1}) and (\ref{eq.sct2}).
Defining the density deviation $\delta\rho(\mathbf{x},t)=\rho(\mathbf{x},t)-\langle \rho(\mathbf{x},t)\rangle$, the definition of the static structure factor, Eq.~(\ref{eq.sf}), can be expressed as
\begin{align}
\begin{split}
	S(\mathbf{k}) = \frac{1}{\langle \rho \rangle} \int \mathrm{d}^2\mathbf{x} \, \mathrm{e}^{-\mathrm{i}\mathbf{k}\cdot\mathbf{x}} \langle \delta\rho(\mathbf{0},t) \delta\rho(\mathbf{x},t) \rangle  \ .
\end{split}
\end{align}
To obtain an explicit expression for $S(\mathbf{k})$,
we linearize Eqs.~(\ref{eq.sct1}) and (\ref{eq.sct2}) around the mean-field solution $(\rho_*,\phi_*)$ given by Eqs.~(\ref{eq.homeostatic.state}) and expect our approximation to be the better the closer the actual expectation values $\langle\rho\rangle$ and $\langle\phi\rangle$ match the mean-field results.
Using the ansatz $\rho=\rho_*+\delta\rho$ and $\phi=\phi_*+\delta\phi$, and expanding in $\delta\rho$ and $\delta\phi$, we obtain the time evolution of the density deviations as
\begin{align}
\begin{split}
	\frac{\partial}{\partial t}\delta\rho
	&\simeq \eta \vec{\nabla}^2 \delta\rho + \alpha\, \delta\phi  +  \sqrt{2\eta \rho_*}\vec{\nabla} \cdot\vec{\xi} + \sqrt{\lambda \rho_*} \zeta  \ , \\
	\frac{\partial}{\partial t}\delta\phi
	&\simeq D \vec{\nabla}^2 \delta\phi - \beta \, \delta\phi - \frac{\Omega^2}{\alpha}\delta\rho  \ ,
\end{split} \label{eq.sct.lin} 
\end{align}
where
\begin{align*}
\begin{split}
	\alpha = \frac{\lambda n \rho_*}{2\phi_0} \ , \quad
	\beta =\bigg(1+\frac{n\mu}{2} \bigg)\kappa \ , \quad
	\Omega =\sqrt{\frac{n \mu\lambda \kappa}{2}} \ .
\end{split}
\end{align*}
Fourier transforming Eqs.~(\ref{eq.sct.lin}) yields
\begin{align}
\begin{split}
	\mathrm{i}{\omega}\widehat{\delta\rho}(\mathbf{k},\omega)
		&= \eta \mathbf{k}^2 \widehat{\delta\rho}(\mathbf{k},\omega) - \alpha\, \widehat{\delta\phi}(\mathbf{k},\omega)  -  \hat{\psi}(\mathbf{k},\omega) \ ,\\
	\mathrm{i}{\omega}\widehat{\delta\phi}(\mathbf{k},\omega)
	&= (D \mathbf{k}^2 + \beta) \widehat{\delta\phi}(\mathbf{k},\omega) + \alpha^{-1}\Omega^2  \widehat{\delta\rho}(\mathbf{k},\omega) \ .
\end{split} \label{eqs.sct.ft}
\end{align}
with the convention $\hat{f}(\mathbf{k},\omega)=\int \mathrm{d}^2\mathbf{x} \, \mathrm{d}t \,\mathrm{e}^{\mathrm{i}(\omega t-\mathbf{k}\cdot\mathbf{x})} f(\mathbf{x},t)$.
Here, $\smash{\hat{\psi}}$ is the Fourier transform of the noise from both the cellular random walk and the birth-death process,%
\begin{widetext}%
\begin{align}
\begin{split}
	\hat{\psi}(\mathbf{k},\omega) &= \sqrt{\rho_*} \int \mathrm{d}^2\mathbf{x}\, \mathrm{d}t \,\mathrm{e}^{\mathrm{i}(\omega t-\mathbf{k}\cdot\mathbf{x})}  \left[\sqrt{2\eta} \vec{\nabla} \cdot\vec{\xi}(\mathbf{x},t) + \sqrt{\lambda} \zeta(\mathbf{x},t) \right] \ ,
\end{split}
\end{align}
which has the correlation function
\begin{align}
\begin{split}
	\langle \hat{\psi}(\mathbf{k},\omega)\hat{\psi}(\mathbf{k}',\omega') \rangle
	&= 
	(2\pi)^3\rho_* (2\eta \mathbf{k}^2 +\lambda )  \delta(\mathbf{k}+\mathbf{k}')\delta(\omega+\omega') \ .
\end{split} \label{eq.noise.correlation}
\end{align}
The solution to Eqs.~(\ref{eqs.sct.ft}) is given by
\begin{align}
\begin{split}
	\widehat{\delta\rho}(\mathbf{k},\omega) &= \frac{-\mathrm{i} \omega+ D\mathbf{k}^2 + \beta}{\Delta(\mathbf{k},\omega)} \hat\psi(\mathbf{k},\omega) \ , \qquad
	\widehat{\delta\phi}(\mathbf{k},\omega) =- \frac{\gamma }{2}\frac{1}{\Delta(\mathbf{k},\omega)} \hat\psi(\mathbf{k},\omega) \ ,
\end{split}
\end{align}
where $\Delta(\mathbf{k},\omega)=(-\mathrm{i} \omega+\eta \mathbf{k}^2)(-\mathrm{i} \omega +D \mathbf{k}^2 + \beta)+\Omega^2$.
\noindent Using Eq.~(\ref{eq.noise.correlation}), the correlators for $\delta\rho$ and $\delta\phi$ are given by
\begin{align}
\begin{split}
	\langle \widehat{\delta\rho}(\mathbf{k},\omega) \widehat{\delta\rho}(\mathbf{k}',\omega') \rangle
	&= 8\pi^3\rho_* (\omega^2+(D \mathbf{k}^2+\beta)^2) \Lambda(\mathbf{k},\omega)\delta(\mathbf{k}+\mathbf{k}')\delta(\omega+\omega') \ , \\
	\langle \widehat{\delta\phi}(\mathbf{k},\omega) \widehat{\delta\phi}(\mathbf{k}',\omega') \rangle
	&= 2\pi^3\rho_*\gamma^2\Lambda(\mathbf{k},\omega)\delta(\mathbf{k}+\mathbf{k}')\delta(\omega+\omega') \ ,
\end{split}
\end{align}
where
\begin{align}
\begin{split}
	\Lambda(\mathbf{k},\omega)
	&= \frac{2\eta \mathbf{k}^2 + \lambda}{(\omega^2+\eta^2 \mathbf{k}^4)(\omega^2+(D \mathbf{k}^2+\beta)^2)-2\Omega^2[\omega^2- \eta\mathbf{k}^2 (D \mathbf{k}^2+\beta)]+\Omega^4} \ .
\end{split}
\end{align}
\noindent%
Defining the function $\hat{S}(\mathbf{k},\omega)$ through $(2\pi)^3 \delta(\mathbf{k}+\mathbf{k}')\delta(\omega+\omega')\hat S(\mathbf{k},\omega)=\langle \widehat{\delta\rho}(\mathbf{k},\omega) \widehat{\delta\rho}(\mathbf{k}',\omega') \rangle$, the result Eq.~(\ref{eq.sf}) is obtained as the backtransform $S(\mathbf{k})=(2\pi)^{-1} \langle \rho \rangle^{-1}  \int\mathrm{d}\omega \, \smash{\hat S}(\mathbf{k},\omega)$.
\end{widetext}
In the long-wavelength limit $\mathbf{k}\to 0$, the structure factor is given by
\begin{align}
\begin{split}
	S(\mathbf{0}) &= \frac{\lambda}{\sqrt{32} \beta\Omega^2} \frac{(\omega^+_\vec{0}-\omega^-_\vec{0})(2 \beta+\omega^+_\vec{0}\omega^-_\vec{0})}{\omega^0_\vec{0}}\ ,
\end{split}
\end{align}
with the functions $\omega^0$ and $\omega^\pm$ given by Eqs.~(\ref{eq.sf.aux}). 
In the active phase ($\mu>0$) and with non-vanishing parameters, this expression remains finite.
The limiting case $n\to 0$, where $n$ is the exponent of the nonlinear feedback function Eq.~(\ref{eq.hill}), corresponds to a random walk with an interactionless critical birth-death process ($p=h(\phi)=1/2$);
since $\Omega \sim \sqrt{n}$, $S(\mathbf{0})$ diverges in the limit $n\to 0$, signaling giant density fluctuations \cite{Ramaswamy2003}. In this case, the structure factor reduces to Eq.~(\ref{sf.cbd}).

An analogous calculation yields the corresponding structure factor $S_\phi$ for the distribution of fate determinants,
\begin{align}
\begin{split}
	S_\phi(\mathbf{k}) &= \frac{1}{\langle \phi \rangle} \int \mathrm{d}^2 \mathbf{x} \, \mathrm{e}^{-\mathrm{i}\mathbf{k}\cdot\mathbf{x}} \langle \delta\phi(\mathbf{0},t) \delta\phi(\mathbf{x},t) \rangle \\
	&= \frac{\rho_0 \gamma^2 (2u_\vec{k}+\lambda)}{\sqrt{32} \phi_0 (\omega^+_\vec{k}+\omega^-_\vec{k})(u_\vec{k}v_\vec{k}+\Omega^2)}  \ ,
\end{split}
\end{align}
where we have used the definitions Eqs.~(\ref{eq.sf.aux}).

\section{Absence of patterning instabilities}
\label{appendix:patterning.instabilities}
\noindent%
We show that the deterministic limit of the continuum theory, i.e., Eqs.~(\ref{eq.sct1}) and (\ref{eq.sct2}) with $\vec{\xi}=0$ and $\zeta=0$, does not exhibit any patterning instabilities.
Linearizing Eqs.~(\ref{eq.sct1}) and (\ref{eq.sct2}) around the homogeneous state $(\rho,\phi)=(\rho_*,\phi_*)$, given by Eqs.~(\ref{eq.homeostatic.state}), and Fourier transforming in space yields
\begin{align}
	\frac{\partial}{\partial t} \bigg(
	\begin{array}{c} \widehat{\delta\rho}(\mathbf{k},t)\\ \widehat{\delta\phi}(\mathbf{k},t) 
	\end{array}\bigg) =\mathbf{J}(\mathbf{k}) \bigg(
	\begin{array}{c} \widehat{\delta\rho}(\mathbf{k},t)\\ \widehat{\delta\phi}(\mathbf{k},t)
	\end{array}\bigg) \ ,
\end{align}
where the hat denotes the spatial Fourier transform, $\hat{f}(\mathbf{k},t)=\int \mathrm{d}^2\mathbf{x} \,\mathrm{e}^{-\mathrm{i}\mathbf{k}\cdot\mathbf{x}} f(\mathbf{x},t)$, and
 the Jacobian is given by
\begin{align}
	\mathbf{J}(\mathbf{k})=\left(\!
	\begin{array}{cc} -\eta \mathbf{k}^2 & \lambda  \kappa n\mu /\gamma \\
	-\gamma/2 & -D\mathbf{k}^2-\beta
	\end{array}\!\right) \ .
\end{align}
The eigenvalues $e_+$ and $e_-$ of $\mathbf{J}(\mathbf{k})$ are given by
\begin{align}
	2e_\pm(\mathbf{k}) = \operatorname{tr}\mathbf{J}(\mathbf{k}) \pm \sqrt{[\operatorname{tr}\mathbf{J}(\mathbf{k})]^2-4\det \mathbf{J}(\mathbf{k})} \ .
\end{align}
Since $\operatorname{tr}\mathbf{J}(\mathbf{k})<0$, a mode $\mathbf{k}$ could only become unstable if the square root is larger than $|{\operatorname{tr}\mathbf{J}(\mathbf{k})}|$. However, since
\begin{align}
	\det \mathbf{J}(\mathbf{k}) = \eta \mathbf{k}^2(D\mathbf{k}^2+\beta)
	+ \frac{\lambda  \kappa n \mu}{2}  > 0 \ ,
\end{align}
this is never the case, so that for $\mu>0$, the homogeneous state is always stable in the active phase.

\section{Quasi-static description of \\ propagating front dynamics}

\label{appendix:quasi.static}
\noindent%
To obtain the effective evolution equation (\ref{eq.fisher}) for the cell density,
we take the noiseless limit of Eq.~(\ref{eq.sct1}), i.e., $\vec{\xi}=0$ and $\zeta=0$.
To further simplify the analysis, we consider a vanishing diffusion $D=0$ of the fate determinant and a Hill exponent $n=1$ for the feedback function $h$.
Non-dimensionalizing the system according to $\rho \to \rho/\rho_*$, $\phi\to\phi/\phi_0$, $\mathbf{x} \to \smash{\sqrt{\lambda/\eta}} \mathbf{x}$ and $t\to \lambda t$, the resulting dynamic equations are given by
\begin{align}
	\frac{\partial\rho}{\partial t} &= \vec{\nabla}^2 \rho + (2{h}(\phi)-1)\rho \ , \label{eq.ct.1} \\[4pt]
	\tau\frac{\partial\phi}{\partial t} &= 1 - \phi + \mu(1- 2{h}(\phi)  \rho) \ , \label{eq.ct.2}
\end{align}
where $\tau=\lambda/\kappa$ is the inertial parameter introduced in Eq.~(\ref{eq.inertia}) and $h$ is given by Eq.~(\ref{eq.hill}). Taking the quasi-static limit $\tau \to 0$  in Eq.~(\ref{eq.ct.2}) and solving for $h$, we obtain
\begin{align}
	h = \frac{1}{2}+\frac{\mu +2+\sqrt{1+\mu  + \mu^2  (1-2 \rho )^2/4}}{2\mu  \rho } \ . \label{eq.mtg.approx}
\end{align}
Using Eq.~(\ref{eq.mtg.approx}) in Eq.~(\ref{eq.ct.1}) yields Eqs.~(\ref{eq.fisher}) and (\ref{eq.fisher.reaction}).

The resulting reaction term $\mathfrak{F}$, given by Eq.~(\ref{eq.fisher.reaction}), is of the Fisher--KPP-type \cite{Cencini2003}, i.e., it satisfies 
(i) $\mathfrak{F}'(0)>0$, which follows from $\mathfrak{F}'(0)=(2 \rho_0)^{-1}$, and
(ii) $\mathfrak{F}'(\rho)<\mathfrak{F}'(0)$ for $0<\rho<1$, which follows from the fact that $\mathfrak{F}''(\rho) = -(1 + \mu)/(\mu^2 (\rho^2 - \rho + \rho_0^2)^{3/2})$ is negative for $0<\rho<1$ and $\mu>0$.

\section{Steady states for two populations with differential responsiveness to the fate determinant}
\label{appendix:advantage}
\noindent%
Here we show that, at the mean-field level, a system with two stem cell populations~`W' and `M' with a differential responsiveness to the fate determinant will lead to extinction of the less responsive population~`W', as detailed in Section~\ref{sec:advantage}.
In the spirit of Eq.~(\ref{eq.reaction.field}), we analyze the steady states of a mean-field system with density coordinate $\vec{Q}=(\rho^\mathrm{W},\rho^\mathrm{M},\phi)$, where $\rho^\mathrm{W}$ and $\rho^\mathrm{M}$ are the mean-field densities of the respective stem cell populations.
Defining the proliferative advantage $a$ of population~`M' via the difference in fate-determinant threshold levels,
\begin{align}
	\phi_0^\mathrm{M}=(1-a)\phi_0^\mathrm{W} \ , \qquad
	0<a<1 \ ,
\end{align}
and rescaling $\phi \to \phi/\phi_0^\mathrm{W}$,
the mean-field system describing the scenario in Section~\ref{sec:advantage} is given by
\begin{align}	
	\frac{\mathrm{d}\vec{Q}}{\mathrm{d}t}= \left( \begin{array}{c}
	 \lambda(2h(\phi)-1)  \rho^\mathrm{W} \\
	 \lambda(2h([1-a]^{-1}\phi)-1)  \rho^\mathrm{M} \\
	 \nu - \kappa \phi - \gamma h(\phi) (\rho^\mathrm{W}+\rho^\mathrm{M})
	\end{array} \right)  \ .
	\label{eq.two.populations}
\end{align}
The first two components clearly reveal that the system can only attain a steady state if either $\rho^\mathrm{W}=0=\rho^\mathrm{M}$ (loss state) or if one of the mutually exclusive conditions (i) $\phi=1$ and $\rho^\mathrm{M}=0$ or (ii) $\phi=1-a$ and $\rho^\mathrm{W}=0$ is fulfilled, showing that both populations cannot stably coexist.
The equilibria with finite cell densities are given by
\begin{align}
	\vec{Q}_*^\mathrm{W} = \left(\!
	\begin{array}{c} \frac{2(\nu-\kappa \phi_0)}{\gamma} \\[3pt]
	0 \\[3pt]
	1
	\end{array}\!\right) \ , \quad
	\vec{Q}_*^\mathrm{M} = \left(\!
	\begin{array}{c} 0 \\[3pt]
	\frac{\nu-\kappa+a\kappa}{\gamma h(1-a)} \\[3pt]
	1-a
	\end{array}\!\right) \ .
\end{align}
Notably, homeostasis of the population~`W' is always unstable: one of the eigenvalues of the Jacobian matrix of the reaction field Eq.~(\ref{eq.two.populations}) evaluated at $\vec{Q}_*^\mathrm{W}$ is given by $\lambda a/(2-a)$, which is always positive for $0<a<1$. This shows that, on the mean-field level, the population~`W' always becomes extinct in the presence of the population~`M'. In contrast, the stability of $\vec{Q}_*^\mathrm{M}$ depends on the kinetic parameters of the fate determinant in a  way similar to the one derived in Appendix~\ref{appendix:stability.analysis} for the one-population homeostatic state.

\section{Non-neutral clonal dynamics \\ in $\boldsymbol{d=1}$ dimension}

\label{appendix:biased.voter.model}
\noindent%
We derive the asymptotic solution of the biased lattice voter model in $d=1$ dimensions. In one dimension, the clone size $s$ (i.e., the number of contiguous lattice sites with the higher proliferation rate) is equivalent to the position of its boundary, whose dynamics is described by a random walk with a directional bias of degree ${v}$ \cite{Snippert2013}. The corresponding master equation for the probability $\Pi=\Pi(s,t)$ to find a clone of size $s$ at time $t$ on a periodic domain of length $\ell$ is given by
\begin{align}
\begin{split}
	\frac{\partial }{\partial t}\Pi &= \bigg\{ (1+{v}) \hat{S}^- - 2 + (1-{v}) \hat{S}^+ \\
	&\qquad - \delta_{s,1} (1+{v}) \hat{S}^- - \delta_{s,\ell-1} (1-{v}) \hat{S}^+ \Big. \\
	&\qquad + 2(\delta_{s,0}+\delta_{s,\ell}) \bigg\} \Pi \ ,
\end{split}
\end{align}
where we have set the total transition rate to unity and where we have implied the boundary conditions $\Pi(-1,t) = 0=\Pi(\ell+1,t)$. Here, $\smash{\hat{S}^\pm}$ are ladder operators defined by $\smash{\hat{S}^\pm}\Pi(s,t) = \Pi(s\pm 1,t)$. The solution away from the boundaries is given by \cite{Snippert2013}
\begin{align}
\begin{split}
	\Pi(s,t)\Big|_{1\leq s \leq \ell-1} &= \frac{2}{\ell}\mathrm{e}^{\varepsilon(s-1)} \sum_{j=1}^{\ell-1} \sin (k_j) \sin (sk_j) \\
	&\hspace{2cm} \times \mathrm{e}^{-2 \sqrt{1-{v}^2} \Sigma(k_i) t } \ ,
\end{split} \label{eq.voter.sol}
\end{align}
where we have defined $k_j = \pi j / \ell$, the composite parameter $\varepsilon$ as in Eq.~(\ref{eq.biased.voter.parameters}), as well as
\begin{align}
	\Sigma(k) = 2 \left(\sin\frac{k}{2}\right)^2 + \frac{1}{\sqrt{1-{v}^2}}-1 \ .
\end{align}
The solutions at the absorbing boundaries $s=0$ and $s=\ell$ have a different form, which we neglect here since we are interested only in the surviving population.
The continuum limit ($\ell\to\infty$) is obtained by replacing $\ell^{-1} \smash{\sum_{j=1}^{\ell-1}} f(k_j) \to \pi^{-1} \smash{\int_0^\pi \mathrm{d}k \, f(k)}$ in Eq.~(\ref{eq.voter.sol}).
A stationary phase approximation, i.e., expanding $\sin(k/2)^2 \simeq k^2/4 + O(k^4)$ in the function $\Sigma(k)$ and carrying out the integral in Eq.~(\ref{eq.voter.sol}), yields the contribution of the dominant modes in the asymptotic limit. The resulting expression contains error functions (erf) of time, which in the asymptotic limit converge to unity. Thus, we arrive at the approximation $\Pi(s,t) \simeq \smash{\tilde{\Pi}}(s,t)$ with
\begin{align}
\begin{split}
	\tilde{\Pi}(s,t) \Big|_{s>0} &= \sqrt{\frac{ \varepsilon}{2 \pi  r  t}} \mathrm{e}^{(\varepsilon^2 +2)\sqrt{1-{v}^2} t -2 t} \\
	&\qquad \times \big( \mathrm{e}^{-\varepsilon (s-1-r t)^2/(2r  t)} \\
	&\qquad\qquad - \mathrm{e}^{-2\varepsilon} \mathrm{e}^{-\varepsilon (s+1-r t)^2/(2r  t)} \big)  \ , 
\end{split}
\end{align}
with $r$ given by Eq.~(\ref{eq.biased.voter.parameters}).
Considering the Gaussian nature of $\smash{\tilde{\Pi}}$, the lower integration bound for any average of the surviving population can be extended to $-\infty$ in the limit of long times, so that the \emph{surviving} clone size distribution $P$ can be approximated by
\begin{align}
	P(s,t) \equiv \frac{\Pi(s,t)}{1-\Pi(0,t)} \simeq \frac{\tilde{\Pi}(s,t)}{\int_{-\infty}^\infty \tilde{\Pi}(s',t) \, \mathrm{d}s'} \ ,
\end{align}
which results in Eq.~(\ref{eq.vm.clone.sizes}).

\section{Quantized stationary states}
\label{appendix:a.quantum.tissue}

\noindent
In this Appendix, we derive the quantization condition Eq.~(\ref{eq.quantization}) from Eq.~(\ref{eq.ischroedinger}) in the limiting case of a infinitely sharp feedback $h(\phi)=\Theta(\phi-\phi_0)$, where $\Theta$ is the Heaviside function.
Considering a symmetric fate-determinant source $J(\theta)=J(-\theta)$ and rescaling time and angles according to $t\to\lambda t$ and $\theta \to \theta/\smash{\sqrt{\tilde \eta/\lambda}}$ where $\tilde\eta$ is the angular motility constant\footnote{The angular motility $\tilde\eta$ is related to the `Euclidean' motility constant $\eta$ by $\tilde\eta=\eta/R^2$ where $R$ is the radius of the periodic domain.}, the density equation (\ref{eq.ischroedinger}) reduces to
\begin{align}
	\frac{\partial\rho}{\partial t} &= \frac{\partial^2\rho}{\partial\theta^2} + \operatorname{sign}(\phi-\phi_0)  \rho \ .
\end{align}
In contrast to a complex quantum wavefunction, the real density field $\rho$ does not oscillate and, hence, the corresponding stationary equation, $0 = \mathrm{d}^2 \rho/\mathrm{d}\theta^2 + \operatorname{sign}(\phi-\phi_0)  \rho$, does not contain an energy eigenvalue that can possibly suffer from quantization; the only free parameters are those that parametrize the real solution.
Motivated by the symmetry of the source, we now assume that the stationary state arranges such that there are two regions for $\phi(\theta)$, one in which $\phi>\phi_0$ (close to the source) and one in which $\phi<\phi_0$ (away from the source), and define $\vartheta$ as the boundary angle of these regions, i.e., the positive value that satisfies $\phi(\vartheta)=\phi_0$. Hence, the stationary equation becomes
\begin{align}
	 \frac{\mathrm{d}^2\rho}{\mathrm{d}\theta^2} = \operatorname{sign}(|\theta|-\vartheta)\rho \ .
\end{align}
This equation has the piecewise solution
\begin{align}
	\rho(\theta) = 
	\begin{cases}
		c_{\mathrm{int}} \cos\theta & |\theta|<\vartheta \\
		c_{\mathrm{ext}} \cosh(\pi-\theta) & |\theta| >\vartheta
	\end{cases} \ ,
\end{align}
where the constants $c_{\mathrm{int}}$ and $c_{\mathrm{ext}}$ are determined by
continuity and differentiability conditions; these yield the relation $c_\mathrm{int}/c_\mathrm{ext} = \cosh(\pi-\vartheta)/\cos \vartheta$ and the quantization condition
\begin{align}
	c_{\mathrm{ext}} [ \sinh(\pi-\vartheta)-\cosh(\pi-\vartheta) \tan(\vartheta) ] = 0 \ . \label{eq.consistency.condition}
\end{align}
Eq.~(\ref{eq.consistency.condition}) can only be fulfilled for $c_{\mathrm{ext}}=0$ or if the term in parentheses vanishes. Rearranging Eq.~(\ref{eq.consistency.condition}) for $c_{\mathrm{ext}}\neq 0$ yields the parameter-free equation~(\ref{eq.quantization}). This derivation shows that the angle $\vartheta$, which separates the self-renewal and differentiation zones, is determined by the dynamics of $\rho$ alone. 
In principle, given the stationary solution $\rho(\theta)$, the corresponding  solution for $\phi$ could be derived from Eq.~(\ref{eq.sct2}). However, even in limiting cases, this yields lengthy expressions which do not provide any further qualitative insights into the quantization mechanism.

\end{appendix}

\end{document}